\documentclass[aps,twocolumn,prd,superscriptaddress,nofootinbib]{revtex4}
\usepackage[dvips]{graphicx}
\usepackage{amsmath}
\usepackage{amssymb}
\newcommand{\beq}{\begin{equation}}
\newcommand{\eeq}{\end{equation}}
\newcommand{\ave}[1]{\langle {#1} \rangle}
\newcommand{\eq}[1]{Eq.~(\ref{#1})}
\newcommand{\eqs}[1]{Eqs.~(\ref{#1})}
\newcommand{\fig}[1]{Fig.~\ref{#1}}
\newcommand{\tab}[1]{Table~\ref{#1}}
\renewcommand{\=}{\;=\;}
\newcommand{\vect}[2]{\left(\!\!\begin{array}{c}{#1} \\ {#2} 
\end{array}\!\!\right)}
\newcommand{\matr}[4]{\left(\!\!\begin{array}{cc}{#1} & {#2} \\ {#3} & {#4} 
\end{array}\!\!\right)}
\def\psl{p\hspace{-1.7mm}/}
\def\pslm{\psl_-}
\def\pslp{\psl_+}
\def\qsl{q\hspace{-2.0mm}/}
\def\dsl{\partial\hspace{-2.0mm}/}
\def\unity{1\hspace{-1.4mm}1}

\def\dfk{\frac{d^4 k}{(2\pi)^4}}
\def\dtk{\frac{d^3 k}{(2\pi)^3}}

\def\intk{\int\!\dfk}
\def\Mintk{T\sum_n\int\dtk}
\def\={\;=\;}
\def\+{\;+\;}
\def\bear{\begin{array}}
\def\ear{\end{array}}

\def\TrNG{\frac{1}{2}\mathrm{Tr}}
\begin{document}

\title{Pseudoscalar Goldstone bosons in the color-flavor locked phase
at moderate densities}

%%%%%%%%%%%%% begin addresses %%%%%%%%%%%%%%%%%%%

\author{Verena Kleinhaus}
\affiliation{Institut f\"ur Kernphysik, Technische Universit\"at Darmstadt, Germany}

\author{Michael Buballa}
\affiliation{Institut f\"ur Kernphysik, Technische Universit\"at Darmstadt, Germany}

\author{Dominik Nickel}
\affiliation{Institut f\"ur Kernphysik, Technische Universit\"at Darmstadt, Germany}

\author{Micaela Oertel}
\affiliation{LUTH,Observatoire de Paris,CNRS, Universit\'e Paris
  Diderot, 5 place Jules Janssen,
  92195 Meudon,France}

%%%%%%%%%%%%% end addresses %%%%%%%%%%%%%%%%%%%%%

\date{\today}

%%%%%%%%%%%%% begin abstract %%%%%%%%%%%%%%%%%%%%

\begin{abstract}

The properties of the pseudoscalar Goldstone bosons in the 
color-flavor locked phase at moderate densities are studied 
within a model of the Nambu--Jona-Lasinio type. 
The Goldstone bosons are constructed explicitly by solving
the Bethe-Salpeter equation for quark-quark scattering in 
random phase approximation.
Main focus of our investigations are 
(i) the weak decay constant in the chiral limit, (ii) the masses of 
the flavored (pseudo-) Goldstone bosons for non-zero but equal quark 
masses, (iii) their masses and effective chemical potentials for 
non-equal quark masses, and (iv) the onset of kaon condensation.   
We compare our results with the predictions of 
the low-energy effective field theory.
The deviations from results obtained in the weak-coupling limit
are discussed in detail. 

\end{abstract}

%%%%%%%%%%%%% end abstract %%%%%%%%%%%%%%%%%%%%%%

\maketitle

%%%%%%%%%%%%%%%%%%%%%%%%%%%%%%%%%%%%%%%%%%%%%%%%%
%%%%%%%%%%%%%%%%%%%%%%%%%%%%%%%%%%%%%%%%%%%%%%%%% 

\section{Introduction}

Much effort has recently been devoted to the study of strongly
interacting matter at nonzero baryon density. In particular
the rich phase structure of color superconducting quark matter 
has attracted much interest. 
(For reviews on color superconductivity see, e.g., 
Refs.~\cite{Rajagopal:2000wf,Alford:2001dt,Schaefer,Rischke:2003mt,
Buballa:2003qv,Ren:2004nn,Huang:2004ik,Shovkovy:2004me}.)
In nature, quark matter phases might be realized in compact stars 
\cite{Ivanenko:1965dg,Itoh:1970uw,Collins:1974ky}.
It is therfore natural to ask whether quark pairing has
interesting phenomenological consequences for compact star physics.
In this context the energetically lowest lying degrees of freedom 
are relevant for many dynamic properties of quark matter.

At low temperatures and very high densities the preferred state is most 
probably the color-flavor locked(CFL) phase where up, down, and strange
quarks are paired in a particularly symmetric way ~\cite{CFL}.
This can be shown from first principles within a weak-coupling 
expansion~\cite{Shovkovy:1999mr,Schafer:1999fe,Evans:1999at}. 
Although this expansion is not valid at ``moderate'' densities which
could be reached in compact stars,
recent Dyson-Schwinger studies indicate that the CFL phase might be
the preferred phase all the way down to the hadronic phase 
\cite{Nickel:2006kc}. 

In the CFL phase, all quark flavors and colors participate in a condensate.
As a consequence, all fermionic modes are gapped and do not appear in
the low-energy excitation spectrum.   
The diquark condensates break the original $U(1)_\mathit{baryon}
\times SU(3)_\mathit{color} \times SU(3)_L \times SU(3)_R$ symmetry of
three-flavor QCD (in the chiral limit) down to a residual $Z_2 \times
SU(3)_\mathit{color+V}$, corresponding to a simultaneous (``locked'')
rotation in color and flavor space. Due to the breaking
of the color symmetry, all eight gluons receive a mass, while the
breaking of baryon number and chiral symmetry leads to the emergence
of one scalar and eight pseudoscalar Goldstone bosons.  In addition,
there is a ninth pseudoscalar Goldstone boson related to the
spontaneous breaking of $U_A(1)$ which is a symmetry of QCD at very
high density \cite{Schafer:2002ty,Rapp:1999qa}.  
In the presence of quark masses chiral symmetry is
broken explicitly and the pseudoscalar Goldstone bosons aquire a mass,
while the scalar Goldstone boson remains massless. Since, with all
quarks being gapped, the
Goldstone bosons are the lowest lying excitations, they play an
important role for the thermodynamic and transport properties of
strongly interacting matter, relevant for compact star phenomenology
(cf., e.g., Refs.~\cite{Shovkovy:2002kv,Manuel:2004iv,Alford:2007rw}).

The symmetry breaking pattern is the basis for the construction of
the low-energy effective theory (LEET)~\cite{Casalbuoni:1999wu,SS99,
Schafer:2000ew,Casalbuoni:2000na,BS2002,Kaplan:2001qk,Yamamoto:2007ah}, 
which describes the Goldstone boson dynamics and is valid for energies 
much smaller than the superconducting gap.
At very high densities, the interaction is weak and
the constants for the LEET can be
calculated from QCD using High Density Effective Theory 
(HDET)~\cite{BS2002,Hong,Beane:2000ms,Nardulli:2002ma}. For
instance, in the weak-coupling limit, pseudoscalar meson masses and
decay constants have been 
investigated~\cite{SS99,Beane:2000ms,RSWZ,Hong:1999ei,Manuel:2000wm,
Miransky:2000bd}.
It was also shown that the stress imposed by the strange quark mass
on the CFL Cooper pairs acts as an effective strangeness chemical 
potential, which may eventually lead to kaon condensation 
\cite{Schafer:2000ew,BS2002,Kaplan:2001qk}.

At intermediate densities, relevant for compact star phenomenology, the
interaction becomes non-perturbative and it is difficult to study the
Goldstone boson dynamics from first principles. The leading-order
predictions, however, are often universal, in the sense that they do not
depend on the interaction, but should hold in any model exhibiting the
same symmetry pattern. One such model is the Nambu--Jona-Lasinio
(NJL) model \cite{NJL}, often used in
the intermediate density regime to study at least qualitatively the
main features. (For reviews, see, e.g., Refs.~\cite{Buballa:2003qv,
Vogl:1991qt,Klevansky:1992qe,Hatsuda:1994pi}.)
This model has already been applied to study kaon condensation in the 
CFL phase at non-zero strange quark masses 
\cite{Buballa:2004sx,Forbes:2004ww,Warringa:2006dk}. 
However, this was done by focusing on the ground state properties,
i.e., without explicit construction of the Goldstone bosons. 
In Refs.~\cite{Fukushima:2005gt,Ebert}, on the other hand, meson and diquark 
properties in the CFL phase have been studied explicitly, but this 
investigation was restricted to the chiral limit. 
(Mesons and diquarks in the 2SC phase have been discussed in 
Refs.~\cite{blaschke2SC,ebert2SC}.)

In the present paper we discuss a detailed analysis of properties of
pseudoscalar mesons\footnote{In this article, we often use the words
``mesons'' and ``diquarks'' synonymously, see Sec.~\ref{mesons} for
more details.}
in the CFL phase in an NJL-type model including
the cases of equal and unequal nonzero quark masses. Emphasis is put on a
comparison with the weak-coupling results.

The paper is organized as follows. In Sec.~\ref{sec:formalism} we
introduce our model and
discuss how to construct the mesonic excitations. Sec.~\ref{eqmass} is
devoted to some general results which can be obtained in the limit of
equal quark masses based on chiral Ward-Takahashi identities. In
Sec.~\ref{numerics} numerical results will be presented. Within
that section, we investigate the pion decay constant in the chiral limit
as well as meson masses for the cases of equal and unequal quark masses.
In this context we also discuss the onset of kaon condensation in the CFL 
phase. Our results are summarized in Sec.~\ref{summary}.

%%%%%%%%%%%%%%%%%%%%%%%%%%%%%%%%%%%%%%%%%%%%%%%%%%%%%%%%%%%%%%% 

\section{Formalism}
\label{sec:formalism}
\subsection{Model Lagrangian}

We consider an NJL-type Lagrangian
\beq
\mathcal{L} = \bar{q} (i \dsl - \hat{m}) q 
+  \mathcal{L}_\mathit{qq},
\label{L}
\eeq
where $q$ is a quark field with three flavor and three color degrees 
of freedom, $\hat m=\mathrm{diag}_f(m_u,m_d,m_s)$ is the mass matrix, and
\begin{alignat}{1}
\mathcal{L}_\mathit{qq}
= H \hspace{-3mm} \sum_{A,A'=2,5,7} \big[ \quad 
                      & ( \bar{q} i \gamma_5 \tau_A \lambda_{A'} C\bar{q}^T) 
                      ( q^T C i \gamma_5 \tau_A \lambda_{A'} q) 
\nonumber\\
            +\;         & ( \bar{q} \tau_A \lambda_{A'} C \bar{q}^T)
                      ( q^T C \tau_A \lambda_{A'} q )\; \big]
\label{Lqq}
\end{alignat}
describes an $SU(3)_\mathit{color} \times U(3)_L\times U(3)_R$
symmetric four-point interaction with a dimensionful coupling constant $H$.
$C=i \gamma^2 \gamma^0$ is the matrix of charge conjugation,
and $\tau$ and $\lambda$, denote Gell-Mann matrices acting in flavor 
space and color space, respectively. 
In this article, we follow the convention that the indices 
$A$ and $A'$ are used for the antisymmetric Gell-Mann matrices only, i.e.,
$A,\, A' \in \{2,5,7\}$,
whereas arbitrary Gell-Mann matrices will be denoted by small Latin letters, 
e.g., $\tau_a$, $a = 1,\dots,8$.

The first term in \eq{Lqq}
corresponds to a scalar quark-quark interaction in the color 
and flavor antitriplet channel, 
just as needed for giving rise to the diquark condensates in the CFL phase
(see \eq{saa} below). 
The second term is the corresponding pseudoscalar interaction and is 
required by chiral symmetry. This will be the essential term for the
pseudoscalar Goldstone excitations we want to study.

For simplicity, we restrict ourselves to quark-quark interactions.
The effect of quark-antiquark interactions, which give rise to normal
self-energies and thereby to dynamical quark masses, will be investigated
in a future publication.

\subsection{Operators in Nambu-Gorkov space}

Introducing Nambu-Gorkov bispinors, 
\beq
    \Psi \= \frac{1}{\sqrt{2}}\,\vect{q}{C\bar{q}^T},
\eeq
\eq{Lqq} can be rewritten as
\begin{alignat}{1}
\mathcal{L}_\mathit{qq} \= 4H\hspace{-3mm}\sum_{A,A'=2,5,7}
\Big\{\quad &\bar\Psi\, \Gamma_{AA'}^{s\uparrow}\,\Psi\;
         \bar\Psi\,\Gamma_{AA'}^{s\downarrow}\,\Psi
\nonumber\\
      \+ &\bar\Psi\,\Gamma_{AA'}^{p\uparrow}\,\Psi\;
         \bar\Psi\,\Gamma_{AA'}^{p\downarrow}\,\Psi\;\Big\},
\end{alignat}
with 18 scalar operators
\beq
\Gamma_{AA'}^{s\uparrow} 
\= \matr{0}{i \gamma_5 \tau_A \lambda_{A'}}{0}{0}, \quad
\Gamma_{AA'}^{s\downarrow}
\=\matr{0}{0}{i \gamma_5 \tau_A \lambda_{A'}}{0}
\label{sop}
\eeq
and 18 pseudoscalar operators
\beq
\Gamma_{AA'}^{p\uparrow} 
\= \matr{0}{\tau_A \lambda_{A'}}{0}{0}, \quad
\Gamma_{AA'}^{p\downarrow}
\=\matr{0}{0}{\tau_A \lambda_{A'}}{0}.
\label{pop}
\eeq
From these expressions we obtain the scattering kernel 
\beq
    \hat K \= \Gamma_i\,K_{ij}\,\bar\Gamma_j, 
\label{kopNG}
\eeq
where $\Gamma_i$ are the 36 operators defined above, 
\beq
    \bar\Gamma_i \= \gamma^0 \Gamma_i^\dagger \gamma^0,
\eeq
and
\beq
    K_{ij} \= 4H\,\delta_{ij}.
\label{kijNG}
\eeq
Repeated operator indices are summed over, unless stated otherwise.

Vertices describing the coupling of an external source to a bare quark
are generalized to Nambu-Gorkov space in the following way:
\beq
    \hat\Gamma \;\rightarrow\; (\hat\Gamma)_\mathit{NG} \;\equiv\;
    \matr{\hat\Gamma}{0}{0}{-C\,\hat\Gamma^T\,C}.
\label{GammaNG}
\eeq
This guarantees that quark-antiquark bilinears remain unchanged,
$\bar\Psi\,(\hat\Gamma)_\mathit{NG}\,\Psi \= \bar q\,\hat\Gamma\,q$.

\subsection{CFL ground state}

Before we construct the mesonic excitations, we have to determine the 
ground state of the system. 
For equal quark masses the CFL phase can
be characterized by the equality of three scalar diquark condensates
in the color and flavor antitriplet channel, 
\beq
s_{22} = s_{55} = s_{77},
\label{cfl}
\eeq
where
\beq
    s_{AA'} \= \ave{\,q^T \,C \gamma_5 \,\tau_A \,\lambda_{A'} \,q\,}~.
\label{saa}
\eeq
In general, these condensates are accompanied by induced color-flavor
sextet condensates. These are, however, small and can be neglected.
If the $SU(3)$-flavor symmetry is explicitly broken by unequal quark
masses, \eq{cfl} does no longer hold exactly, but the three condensates
may differ from each other.

To obtain the ground state we must minimize the
thermodynamic potential (per volume $V$),
\beq
    \Omega(T,\{\mu_i\}) \= -\frac{T}{V}\,\ln{\cal Z}(T,\{\mu_i\}),
\eeq
where ${\cal Z}(T,\{\mu_i\})$ is the grand partition function
at temperature $T$ and a given set of chemical potentials $\{\mu_i\}$. 
For $\beta$ equilibrated matter, these can be expressed in terms of
the quark number chemical potential $\mu$, the electric charge chemical
potential $\mu_Q$, and two color chemical potentials $\mu_3$ and $\mu_8$
\cite{SRP}.

For the CFL phase in mean-field approximation, $\Omega$ is given by
\begin{alignat}{1}
\Omega(T,\{\mu_i\})\= 
&-T \sum_{n} \int \frac{d^3p}{(2\pi)^3}  
\frac{1}{2}{\rm Tr}\ln\left({\frac{1}{T}S^{-1}(i\omega_n,\vec p\,)}\right)
\nonumber\\
&+
\frac{1}{4H} \sum_{A=2,5,7}|\Delta_A|^2,  
\label{Omega}
\end{alignat}
where the gap parameters $\Delta_A$ are related to the diquark condensates,
\beq
    \Delta_A \= -2H\,s_{AA}.
\eeq
The inverse dressed propagator reads
\beq
S^{-1}(p) \= \matr{\psl + \hat \mu\gamma^0 - \hat m}
             {\sum\limits_{A=2,5,7}\Delta_A\gamma_5\tau_A\lambda_A}
      {-\hspace{-3mm}\sum\limits_{A=2,5,7}\Delta_A^*\gamma_5\tau_A\lambda_A}
             {\psl - \hat \mu\gamma^0 - \hat m}.
\label{SinvNG}
\eeq
Here $\hat\mu$ denotes the diagonal matrix in color and flavor space
which is given by the set of chemical potentials $\{\mu_i\}$.

In order to determine the ground state, $\Omega$ must be minimized with
respect to the gap parameters $\Delta_A$, leading to three gap equations
\beq
    \frac{\partial\Omega}{\partial\Delta_A^*} \= 0, \quad A = 2,5,7.
\label{gapOm}
\eeq 
Furthermore, we require the solutions to be color and electrically 
neutral in the presence of leptons. In general, this leads to three 
additional equations,
\beq
    n_i \equiv -\frac{\partial\Omega_\mathit{tot}}{\partial\mu_i} \= 0, 
    \quad i = Q,3,8,
\eeq 
where $\Omega_\mathit{tot} = \Omega + \Omega_\mathit{leptons}$ is the
sum of the quark part, \eq{Omega}, and the contribution of the leptons.
Thus, altogether we have a set of six coupled equations for $\Delta_A$
and $\mu_i$ which must be solved simultaneously. 
This has been done many times before, and we can refer to the literature
for technical details, e.g., 
Refs.~\cite{SRP,Ruster:2005jc,Blaschke:2005uj,Abuki:2005ms}.
In the present article, we restrict ourselves to the (fully gapped) CFL 
phase at zero temperature. In this case, color neutral quark matter
is electrically neutral without leptons \cite{Rajagopal:2000ff}
and $\mu_Q = 0$. Moreover, we consider isospin symmetry, $m_u = m_d$,
so that $\mu_3$ vanishes as well.

%%%%%%%%%%%%%%%%%%%%%%%%%%%%%%%%%%%%%%%%%%%%%%%%%%%%%%%%%%

\begin{figure}
\begin{center}
 \includegraphics[width=\linewidth]{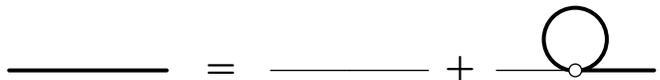}
 \caption{Dyson equation for the dressed Nambu-Gorkov quark 
          propagator (thick line). 
          The thin line indicates to the bare propagator.}
\label{fig:dyson_eq}
\end{center}
\end{figure}

%%%%%%%%%%%%%%%%%%%%%%%%%%%%%%%%%%%%%%%%%%%%%%%%%%%%%%%%%%%%

The gap equations (\ref{gapOm}) can be derived from the Dyson equation 
for the dressed quark propagator, too, diagrammatically shown in 
Fig.~\ref{fig:dyson_eq}. 
This is well known, but some details are useful in our later discussion. 
Therefore, we present this derivation in Appendix~\ref{appgap}.

\subsection{Axial transformations}

In the chiral limit, $m_u=m_d=m_s=0$, the Lagrangian, \eq{L}, is
invariant under $SU(3)_\mathit{color}\times U(3)_L\times U(3)_R$
transformations.  In the CFL phase, this symmetry is spontaneously
broken to the diagonal vector subgroup, $SU(3)_{\mathit{color}+V}$.
This is reflected by nonzero values for the condensates, \eq{cfl}.
Axial transformations,
%%%%%%%%%%%%%%%%%%%%%%%%%%%%%%%%%%%%
\beq q \;\rightarrow\; q' \=
\exp(i\theta_a\,\gamma_5\,t_a)\,q,
\label{axialq}
\eeq 
%%%%%%%%%%%%%%%%%%%%%%%%%%%%%%%%%%%%%5
then connect a
continuous set of degenerate ground states for the CFL phase. 
Here $t_a = \frac{\tau_a}{2}$, $a=0,\dots,8$, where $\tau_1,
\dots, \tau_8$ denote the eight Gell-Mann matrices in flavor space, as
before, and $\tau_0 = \sqrt{2/3}\,\unity_f$. 
These 
transformations can be parameterized by nine pseudoscalar Goldstone 
bosons. In order to identify the operators corresponding to Goldstone
boson excitations, we inspect the effect of an axial
transformation with specified quantum numbers on the condensates, \eq{cfl}. 
In Nambu-Gorkov formalism this can be
described as follows
%%%%%%%%%%%%%%%%%%%%%%
\beq 
\ave{\bar\Psi \Gamma_i\Psi}
\;\rightarrow\; \ave{\bar\Psi' \Gamma_i\Psi'} \;\equiv\; \ave{\bar\Psi
  \Gamma_i'\Psi}, 
\eeq 
%%%%%%%%%%%%%%%%%%%%%%%
for $\Gamma_i = \Gamma_{AA}^{s\uparrow}$ and
$\Gamma_{AA}^{s\downarrow}$.  Considering infinitesimal
transformations, we find 
%%%%%%%%%%%%%%%%%%%%%%%%
\beq 
\Gamma_i' \= \Gamma_i \+ i\theta_a
\delta\Gamma_{i,a}, \eeq with \beq \delta\Gamma_{i,a} \=
\Big\{(\gamma_5\,t_a)_{NG} \,,\, \Gamma_i \Big\},
\label{deltaGamma}
\eeq
%%%%%%%%%%%%%%%%%%%%%%
where $\{A,B\} = AB + BA$ denotes the anticommutator.

The Goldstone bosons are coupled to the quarks by vertices related to
$\delta\Gamma_{i,a}$ (see Ref.~\cite{RSWZ} for a detailed discussion).  
Thus, by evaluating \eq{deltaGamma} for 
$\Gamma_i = \Gamma_{AA}^{s\uparrow}$ and $\Gamma_{AA}^{s\downarrow}$,
we can determine the operators
which contribute to the vertex of a given Goldstone boson.
It turns out that the assignment is most simple 
if we work in the ``particle basis'' instead of using 
hermitian flavor operators. In this case all flavored mesons
are coupled to only two Nambu-Gorkov operators, 
while the hidden-flavor mesons are coupled to four ($\pi^0$)
or six ($\eta_8$,$\eta_0$) operators.
The results are summarized in Table~\ref{tab:qnop1}.
As we will see below, this assignment remains basically unchanged
when we include finite quark masses. 
%%%%%%%%%%%%%%%%%%%%%%%%%%%%%%%%%%%%%%%%%%%%%%%%%%%%%%%%%%%%%%%%%%%%%
\begin{table}
\begin{center}
\begin{tabular}{c c c}
%\hline
&&\\[-3mm]
\;``meson''\; & $t_j$ & $\Gamma_j$
\\[1mm]
\hline
&&\\[-3mm]
 $\pi^+$ & \quad$\frac{\tau_1 + i\tau_2}{2\sqrt{2}}$\quad
& \quad$\Gamma_{57}^{p\uparrow}$,\; $\Gamma_{75}^{p\downarrow}$\quad 
\\[1mm]
 $\pi^-$ & \quad$\frac{\tau_1 - i\tau_2}{2\sqrt{2}}$\quad 
& \quad$\Gamma_{75}^{p\uparrow}$,\; $\Gamma_{57}^{p\downarrow}$\quad
\\[1mm]
 $K^+$ & \quad$\frac{\tau_4 + i\tau_5}{2\sqrt{2}}$\quad 
& \quad$\Gamma_{27}^{p\uparrow}$,\; $\Gamma_{72}^{p\downarrow}$\quad
\\[1mm]
 $K^-$ & \quad$\frac{\tau_4 - i\tau_5}{2\sqrt{2}}$\quad
& \quad$\Gamma_{72}^{p\uparrow}$,\; $\Gamma_{27}^{p\downarrow}$\quad
\\[1mm]
 $K^0$ & \quad$\frac{\tau_6 + i\tau_7}{2\sqrt{2}}$\quad
& \quad$\Gamma_{25}^{p\uparrow}$,\; $\Gamma_{52}^{p\downarrow}$\quad
\\[1mm]
 $\bar{K}^0$ & \quad$\frac{\tau_6 - i\tau_7}{2\sqrt{2}}$\quad
& \quad$\Gamma_{52}^{p\uparrow}$,\; $\Gamma_{25}^{p\downarrow}$\quad
\\[1mm]
 $\pi^0$ & \quad$\frac{\tau_3}{2}$\quad
& \quad$\Gamma_{55}^{p\uparrow}$,\; $\Gamma_{77}^{p\uparrow}$,\;
$\Gamma_{55}^{p\downarrow}$,\;$\Gamma_{77}^{p\downarrow}$\quad
\\[1mm]
$\begin{array}{c} \eta_8 \\ \eta_0 \end{array}$ 
& \quad$\left.
\begin{array}{c}
\frac{\tau_8}{2} \\ [1mm]
\frac{\tau_0}{2}
\end{array}\right\}$\quad
&
\quad$\left\{\begin{array}{c}
\Gamma_{22}^{p\uparrow},\; \Gamma_{55}^{p\uparrow}, \; \Gamma_{77}^{p\uparrow},
\\[1mm]
\Gamma_{22}^{p\downarrow}, \; \Gamma_{55}^{p\downarrow},\; \Gamma_{77}^{p\downarrow}
\end{array}\right\}$
\end{tabular}
\end{center}
\caption{Pseudoscalar Goldstone modes, corresponding flavor operators $t_j$,
and Nambu-Gorkov operators $\Gamma_j$ contributing to the quark-meson vertex.
The $\Gamma_j$ are obtained by evaluating \eq{deltaGamma} for $t_a=t_j$ and
all possible $\Gamma_i = \Gamma_{AA}^{s\uparrow}$ or
$\Gamma_{AA}^{s\downarrow}$.}
\label{tab:qnop1}
\end{table}  
%%%%%%%%%%%%%%%%%%%%%%%%%%%%%%%%%%%%%%%%%%%%%%%%%%%%%%%%%%%%%%%%%%%%%

\subsection{Mesonic excitations}
\label{mesons}

\begin{figure}
\begin{center}
 \includegraphics[width=\linewidth]{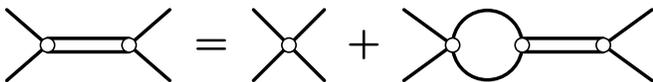}
 \caption{RPA equation for the T-matrix in Nambu-Gorkov space.}
\label{fig:T-matrix}
\end{center}
\end{figure}

Iterating the scattering kernel as illustrated in \fig{fig:T-matrix}
leads to the RPA equation for the T-matrix in Nambu-Gorkov space, 
\beq
\hat T \= \hat K \+ \hat K \hat J~\hat T.
\label{BSETNG}
\eeq Since the interaction is separable, this operator equation can be
converted into a matrix equation by using \eq{kopNG} for the
scattering kernel $\hat K$, an analogous expression for $\hat T$, 
\beq
\hat T \= \Gamma_i\,T_{ij}\,\bar\Gamma_j, \eeq 
and defining 
\beq
\bar\Gamma_i\,\hat J\,\Gamma_j \= J_{ij}.  
\eeq 
One finds 
\beq T \= K
\+ K J~T, 
\eeq 
with the solution 
\beq 
T(q) \= \left(\unity - K
  J(q)\right)^{-1}\,K \= \left(\frac{1}{4H} - J(q)\right)^{-1}, 
\eeq
where the last equality follows from \eq{kijNG}.  The polarization
matrix, corresponding to the loop in \fig{fig:T-matrix}, is given by
\beq 
  J_{ij}(q) \= i \intk \, \TrNG\left[\bar\Gamma_i S(k+q) \Gamma_j
  S(k) \right].  
\label{Jij}
\eeq 
Here we have introduced a ``vacuum-like''
notation for brevity.  In medium we should replace 
\beq q
\;\rightarrow\; \vect{i\omega_m}{\vec q}, \quad k \;\rightarrow\;
\vect{i\omega_n}{\vec k}, \eeq and \beq i\intk \;\rightarrow\; -\Mintk
\eeq 
with bosonic Matsubara frequencies $i\omega_m$ and fermionic
Matsubara frequencies $i\omega_n$. In the end, the result should be
analytically continued to real external energies. We will use this
notation throughout this paper. For our numerical results, we will
introduce a 3-momentum cutoff $\Lambda$ to regularize the divergent integrals.

The matrices $T$ and $J$ are $36 \times 36$ matrices in operator 
space, corresponding to the 36 operators defined in \eqs{sop} and
(\ref{pop}). It turns out, however, that $J$ and, thus, $T$ are
block diagonal, i.e., only certain combinations of operators occur. 
More precisely, scalar operators do not mix with pseudoscalar ones, 
and each of the resulting $18 \times 18$ blocks can be decomposed further 
into six $2 \times 2$ blocks and one $6 \times 6$ block. 
These blocks carry different quantum numbers and reflect the assignment
to different meson modes, as given in Table~\ref{tab:qnop1}.\footnote{In 
Table~\ref{tab:qnop1} we have listed the pseudoscalar mesons only,
but the scalar sector is completely analogous.}
In particular, the $2 \times 2$ blocks correspond to the flavored mesons.
The $6 \times 6$ blocks, on the other hand, describe the hidden-flavor 
mesons, i.e., $\pi^0$, $\eta_8$ and $\eta_0$ in the pseudoscalar case. 
As in vacuum, these states are mixed for unequal quark masses.  

\begin{figure}
\begin{center}
 \includegraphics[width=0.5\linewidth]{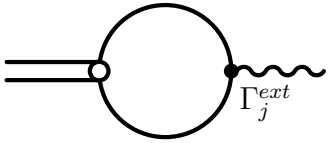}
 \caption{Coupling the T-matrix (double line) to an external meson 
          source (wavy line).}
\label{fig:source-loop-T}
\end{center}
\end{figure}

This assignment can be tested
by coupling the T-matrix to an external meson source,
as illustrated in Fig.~\ref{fig:source-loop-T}.
To that end, we evaluate the loop integral
\beq
    I_{ij}(q) \=
    i\intk\, \TrNG\left[\bar\Gamma_i\,  S(k+q)\,
    \Big(\Gamma^{ext}_j\Big)_\mathit{NG} \,S(k)\right].
\label{Iij}
\eeq
where $\Gamma^{ext}_j$ denotes the vertex of the external source 
to the quark. It is given by $\Gamma^{ext}_j = t_j$ 
for scalar sources and $\Gamma^{ext}_j = i\gamma_5\,t_j$
for pseudoscalar sources. Here $t_j$ is one of the generators in flavor 
space listed in Table~\ref{tab:qnop1}, and
$(\Gamma^{ext}_j)_\mathit{NG}$ is the generalization of this vertex to
Nambu-Gorkov space, as defined in \eq{GammaNG}.

The blocks can be diagonalized by unitary transformations,
\beq
    J' \= W\,J\,W^\dagger,
\eeq
with $W$ being an (in general 4-momentum dependent) unitary matrix and
\beq
    J'_{ij} \= J^{(i)}\,\delta_{ij}.
\eeq
Defining new operators
\beq
    \Gamma_j' = \Gamma_i\,W^\dagger_{ij}, 
\eeq
$J'$ can be rewritten as
\beq
    J_{ij}'(q) \= i \intk \, \TrNG\left[\bar\Gamma_i'  S(k+q) \Gamma_j'  S(k)
    \right].
\eeq
Since the scattering kernel remains diagonal in this new basis,
\beq
    \hat K \= 4H\,\Gamma_i\,\bar\Gamma_i\= 4H\,\Gamma_i'\,\bar\Gamma_i',
\eeq
the T-matrix becomes diagonal as well
\beq
    \hat T \= \Gamma_i'\,T^{(i)}\,\bar\Gamma_i',
\label{TdiagNG}
\eeq
with
\beq
    T^{(i)}(q) \= \frac{1}{\frac{1}{4H} - J^{(i)}(q)}.
\eeq
In the vicinity of a pole, we can parameterize these modes
like a free boson with mass $m_i$ in the presence of a 
chemical potential $\mu_i$ corresponding to this particular boson,
\beq
    T^{(i)}(q) \;\approx\; \frac{-g_i^2}
    {(q_0+\mu_i)^2 - c_i^2\vec q\,^2 - m_i^2}.
\label{TMgen}
\eeq
Here $g_i$ can be interpreted as a coupling constant of the boson to
an external quark, and $c_i$ denotes the in-medium group velocity.
In this article we restrict ourselves to $\vec q = 0$ 
in order to keep artifacts of the 3-momentum cutoff as small as 
possible. 

The various modes $T^{(i)}$ describe bosonic excitations of the CFL ground 
state. Because of the formal quark-antiquark structure of the 
polarization loops, we will call these excitations ``mesons''.
However, it should be kept in mind that the propagators and vertices
entering the polarization loops live in Nambu-Gorkov space and therefore
in principle describe quark-antiquark as well as diquark and antidiquark 
(or di-hole)
excitations. In vacuum or in a normal-conducting medium, these are
independent modes, protected by the conserved baryon number. 
In the CFL phase, however, baryon number
is broken, and quark-antiquark, diquark, and antidiquark states can mix.
As our model Lagrangian does not contain quark-antiquark
interactions, our ``mesons'' are in fact superpositions of
diquarks and antidiquarks or, more important, di-holes. 

In vacuum we have nine scalar and nine pseudoscalar diquarks 
and  nine scalar and nine pseudoscalar antidiquarks.
Since the total
number of states does not change when the states are mixed, there must be
36 meson states in the CFL phase, 18 scalars and 18 pseudoscalars.
(If we included quark-antiquark interactions in our model, we would
obtain 27 scalars and 27 pseudoscalars.)
As we will see below, nine pseudoscalars are massless in the chiral limit, 
while the others stay heavy. This has been found in Ref.~\cite{Ebert}, too, 
within a similar model.

\subsection{Pseudoscalar meson decay constants}

As in vacuum, the pseudoscalar mesons can decay 
weakly. The decay amplitudes are related to the loop integrals
\beq
    F_{ij}'^{\,\mu}(q) \=
    i\intk\, \TrNG\left[\bar\Gamma_i'\,  S(k+q)\,
    (\gamma^\mu\gamma_5\,t_j)_\mathit{NG} S(k)\right],
\label{Fijp}
\eeq
describing the coupling of the meson $i$ (i.e., the one which corresponds 
to the operator $\Gamma_i'$) to an external axial current 
$A^\mu_{5j}$.\footnote{Strictly speaking, \eq{Fijp} describes the 
time-reversed process, i.e., the production of a meson by an incoming axial 
current. 
This choice was made for later convenience in Sec.~\ref{sect:CWTI}.
There we apply chiral Ward identities, which are formulated for incoming  
axial currents, see \eq{WTING}.}
In the following, we are mostly interested in flavored mesons, which
are the main focus of our studies. 
Each flavored meson $i$ couples to only one $t_j$,
as listed in \tab{tab:qnop1}.\footnote{Note that the opposite is not true: 
Since we have 18 pseudoscalars, there are in general two meson states 
$i$ which couple to a given flavor operator $t_j$.}
For simplicity, we denote the flavor operator which fits to the
meson $i$ by $t_i$.

The meson decay constants are related to the on-shell values of these
amplitudes,
\beq
    f_i\,q^\mu \= i g_i F_{ii}'^{\,\mu}(q)\Big|_\mathit{on-shell}\quad,
\eeq
with no summation over the index $i$ on the r.h.s. 
The coupling constant $g_i$ has been defined implicitly in \eq{TMgen}.

In general, there are different values for the time-like ($\mu=0$) 
and the space-like ($\mu = 1,2,3$) decay constants, which differ by the
group velocity of the Goldstone modes.
However, since we only consider mesons with vanishing 3-momenta
in this article, we are restricted to the time-like decay constants.

%%%%%%%%%%%%%%%%%%%%%%%%%%%%%%%%%%%%%%%%%%%%%%%%%%%%%%%%%%%%%%% 

\section{Equal quark masses}
\label{eqmass}

It is rather instructive to investigate the simplified case of equal 
quark masses, $m_u = m_d = m_s \equiv m$. 
In this case, we have only one gap parameter
$\Delta_2 = \Delta_5 = \Delta_7 \equiv \Delta$ and
one common chemical potential $\mu$ in electrically and color neutral 
CFL matter.  
Moreover, the set of the 18 pseudoscalar meson states
consists of two $SU(3)$ octets and two singlets, 
with all mesons in the same multiplet being degenerate.\footnote{The degeneracy
in the octets is due to the residual $SU(3)_\mathit{color+V}$
symmetry of the CFL phase. Thus, even though we start from a $U(3)$
invariant Lagrangian in our model, there is no unbroken symmetry which 
relates the singlet states to the octets.}  
Finally, there is no stress caused by quark mass or chemical potential 
differences, which could act as an effective meson chemical potential
as in \eq{TMgen}. 
Hence, for vanishing 3-momenta,
the pole approximation for the T-matrix, \eq{TMgen}, becomes
\beq
    T^{(i)}(q_0) \;\approx\; \frac{-g_i^2}
    {q_0^2 - m_i^2}.
    \label{Tq0}
\eeq
This yields for the decay constants
\beq
    f_i\,q_0 \= i g_i F_{ii}'^{\,0}(q)\Big|_{q_0 = m_i, \vec q = 0}.
\label{fieqm}
\eeq
We will often refer to the states of the lowest
pseudoscalar octet as ``pions'' and denote their masses, couplings,
and decay constants by $m_\pi$, $g_\pi$, and $f_\pi$, respectively.

\subsection{Dressed vertex functions}

\begin{figure}
\begin{center}
 \includegraphics[width=\linewidth]{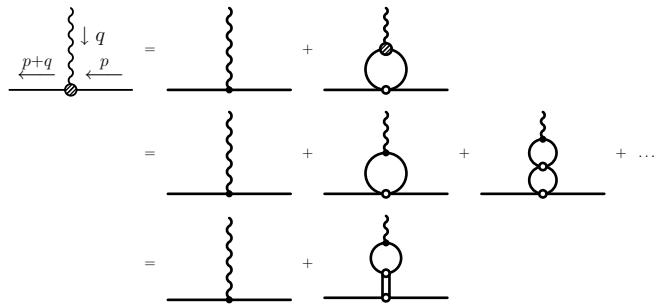}
 \caption{Vertex function for an external axial current (wavy line)
          coupled to a quark. The shaded circles (black dots)
          indicate dressed (bare) vertices. The open circles correspond
          to quark-quark vertices and the double line
          to the T-matrix, as in \fig{fig:T-matrix}.}
\label{fig:dressed_vertex}
\end{center}
\end{figure}

Attaching an external axial current to both sides of the Dyson 
equation for the dressed quark propagator (see upper line of
\fig{fig:dressed_vertex}), 
we obtain a selfconsistency equation for the dressed axial vertex,
\begin{alignat}{1}
    &\Gamma^\mu_{5j}(p;q) \= (\gamma^\mu\gamma_5\,t_j)_\mathit{NG}
\nonumber\\
    &\+ 4iH\;\Gamma_i \intk\, \TrNG\left[\bar\Gamma_i\,  S(k+q)\,
    \Gamma^\mu_{5j}(k;q)\,S(k)\right].
\label{BSEANG}
\end{alignat}
The first term on the r.h.s. corresponds to the bare 
vertex, while the second term contains the dressed vertex again.  
Thus, iterating the equation, the dressed vertex can be written as a Born 
series of quark-antiquark loops (second line of \fig{fig:dressed_vertex}), 
with the bare vertex attached to the last loop. 
Employing the quark-antiquark T-matrix, this could be rewritten
(last line of \fig{fig:dressed_vertex}) as
\beq
    \Gamma^\mu_{5j}(p;q) \=
    (\gamma^\mu\gamma_5\,t_j)_\mathit{NG} 
    \+ T^{(i)}(q)\,\Gamma_i'\,F_{ij}'^{\,\mu}(q),
\eeq
with $F_{ij}'^{\,\mu}$ as defined in \eq{Fijp}.
We have seen earlier that a given flavor operator $t_j$ couples to 
only two meson modes $i$ in flavored channels.
We may call these modes ``$\pi_j$'' and ``$\pi_j'$''.
Then, for $\vec q = 0$ and $q_0$ approaching $m_\pi$, we obtain
\beq
    \Gamma^0_{5j}(p;q_0\rightarrow m_\pi) \= 
     i\frac{g_\pi f_\pi\, q_0}{q_0^2 - m_\pi^2} \,\Gamma_{\pi_j}'
     \+ \text{non-singular terms}.
\label{polax}
\eeq

Similarly, we have an equation for the dressed pseudoscalar vertex
\begin{alignat}{1}
    &\Gamma_{5j}(p;q) \= (i\gamma_5\,t_j)_\mathit{NG}
\nonumber\\
    &\+ 4iH\;\Gamma_i \intk\, \TrNG\left[\bar\Gamma_i\,  S(k+q)\,
    \Gamma_{5j}(k;q)\,S(k)\right],
\label{BSEPNG}
\end{alignat}
which could be rewritten as
\beq
    \Gamma_{5j}(p;q) \=
    (i\gamma_5\,t_j)_\mathit{NG} 
    \+ T^{(i)}(q)\,\Gamma_i'\,I_{ij}'(q),
\eeq
with
\beq
    I_{ij}'(q) \=
    i\intk\, \TrNG\left[\bar\Gamma_i'\,  S(k+q)\,
    (i\gamma_5\,t_j)_\mathit{NG} S(k)\right].
\label{Iijp}
\eeq
Thus, approaching the pole, we find
\beq
    \Gamma_{5j}(p;q_0\rightarrow m_\pi) \,=\, 
     -\frac{g_\pi^2\,I^{(\pi)}}{q_0^2 - m_\pi^2}\,\Gamma_{\pi_j}'
     \,+\, \text{non-singular terms},
\label{polp}
\eeq
with
\beq
    I^{(\pi)} \= I_{\pi\pi}'(q_0^2=m_\pi^2).
\eeq

\subsection{Chiral Ward-Takahashi identity}
\label{sect:CWTI}

It can be shown on general grounds that the exact vertex functions 
and inverse propagators must satisfy the following axial Ward-Takahashi 
identity:
\begin{alignat}{1}
    &q_\mu\,\Gamma^\mu_{5j}(p;q) \+ 2mi\,\Gamma_{5,j}(p;q)
\nonumber\\
    \,=\;
    &S^{-1}(p+q)\,(\gamma_5\,t_j)_\mathit{NG} \,+\, 
    (\gamma_5\,t_j)_\mathit{NG}\,S^{-1}(p).
\label{WTING}
\end{alignat}
In Appendix \ref{appwti} we show by using the gap equation that the 
vertex functions defined above and the inverse propagator 
(\ref{SinvNG}) are consistent with this relation.

We now evaluate this equation for $\vec q = 0$ and 
$q_0 \rightarrow m_\pi$.
In the chiral limit, $m = 0$, only the axial vertex contributes, 
and we find from \eq{polax} that the l.h.s. is given by
\beq
    q_\mu\, \Gamma^\mu_{5j} \= 
     i\frac{q_0^2}{q_0^2 - m_\pi^2} \,g_\pi f_\pi \,\Gamma_{\pi_j}'
     \+ \text{non-singular terms}.
\eeq
In general, this has a singularity at $q_0 = m_\pi$. 
On the other hand, one can easily see that the r.h.s. of \eq{WTING}
remains finite. We thus conclude that the singularity on the l.h.s.
must be suppressed, i.e., either $m_\pi = 0$ or
$g_\pi f_\pi = 0$. In fact, from the symmetry breaking pattern we 
expect nine pseudoscalar Goldstone bosons.
Therefore, both scenarios should be realized, i.e., nine pseudoscalar
mesons (one octet and one singlet) are massless and the others are massive.

For the massless solution, we can now evaluate \eq{WTING} directly 
at $q = 0$. This yields
\beq
     i\,g_\pi f_\pi \,\Gamma_{\pi_j}'\hspace{-2.0mm}^{(0)} \=
     \Big\{ (\gamma_5\,t_j)_\mathit{NG} \,,\, S^{-1}(p) \Big\}.
\label{WTING0}
\eeq
Inserting \eq{SinvNG} for the inverse propagator with $m = 0$
and taking the freedom
to choose the gap parameter $\Delta$ to be real, we find the solution
\beq
    g_\pi\,f_\pi \= \Delta
\label{GTNG}
\eeq
with the assignment (cf. \eq{deltaGamma})
\beq
    \Gamma_{\pi_j}'\hspace{-2.0mm}^{(0)} \= -\!\!\sum\limits_{A=2,5,7}
    \Big\{ (\gamma_5\,t_j)_\mathit{NG} \,,\, 
     (\Gamma_{AA}^{s\uparrow} - \Gamma_{AA}^{s\downarrow}) \Big\}.
\label{Gammaap}
\eeq
\eq{GTNG} may be viewed as a generalization of the well-known
Goldberger-Treiman relation in vacuum~\cite{RSWZ}.

Explicit evaluation of \eq{Gammaap} yields
\begin{alignat}{2}
\Gamma_{\pi^+}'\hspace{-2.5mm}^{(0)} 
&\= \frac{-i}{\sqrt{2}} \Big(\Gamma_{57}^{p\uparrow} 
                              - \Gamma_{75}^{p\downarrow}\Big),\quad &
\Gamma_{\pi^-}'\hspace{-2.5mm}^{(0)} 
&\= \frac{-i}{\sqrt{2}} \Big(\Gamma_{75}^{p\uparrow} 
                              - \Gamma_{57}^{p\downarrow}\Big),
\nonumber\\
\Gamma_{K^+}'\hspace{-2.5mm}^{(0)} 
&\= \frac{i}{\sqrt{2}} \Big(\Gamma_{27}^{p\uparrow} 
                              - \Gamma_{72}^{p\downarrow}\Big),\quad &
\Gamma_{K^-}'\hspace{-2.5mm}^{(0)} 
&\= \frac{i}{\sqrt{2}} \Big(\Gamma_{72}^{p\uparrow} 
                              - \Gamma_{27}^{p\downarrow}\Big),
\nonumber\\
\Gamma_{K^0}'\hspace{-2.5mm}^{(0)} 
&\= \frac{-i}{\sqrt{2}} \Big(\Gamma_{25}^{p\uparrow} 
                              - \Gamma_{52}^{p\downarrow}\Big),\quad &
\Gamma_{\bar{K}^0}'\hspace{-2.5mm}^{(0)} 
&\= \frac{-i}{\sqrt{2}} \Big(\Gamma_{52}^{p\uparrow} 
                              -
                              \Gamma_{25}^{p\downarrow}\Big),
\label{eqmops}
\end{alignat}
for the flavored mesons, and
\begin{alignat}{1}
\Gamma_{\pi^0}'\hspace{-2.0mm}^{(0)}
&\= -\frac{i}{2}\Big( \Gamma_{55}^{p\uparrow} -
\Gamma_{77}^{p\uparrow} - \Gamma_{55}^{p\downarrow} +
\Gamma_{77}^{p\downarrow} \Big), 
\nonumber \\
\Gamma_{\eta_8}'\hspace{-2.0mm}^{(0)} 
&\= \frac{-i}{2\sqrt{3}}\Big( 2\,\Gamma_{22}^{p\uparrow} -\Gamma_{55}^{p\uparrow} -
\Gamma_{77}^{p\uparrow} -  2\,\Gamma_{22}^{p\downarrow} + \Gamma_{55}^{p\downarrow} +
\Gamma_{77}^{p\downarrow} \Big),
\nonumber \\
\Gamma_{\eta_0}'\hspace{-2.0mm}^{(0)} 
&\= -i\sqrt{\frac{2}{3}} \Big( \Gamma_{22}^{p\uparrow} +\Gamma_{55}^{p\uparrow} +
\Gamma_{77}^{p\uparrow} - \Gamma_{22}^{p\downarrow}
-\Gamma_{55}^{p\downarrow} - 
\Gamma_{77}^{p\downarrow} \Big),
\label{eqmopshf}
\end{alignat}
for the mesons with hidden flavor.
Thus, antiparticle modes are related to each other as
\beq
    \Gamma_{\bar{\pi}_j}'\hspace{-1.0mm}^{(0)} \= 
    \bar\Gamma_{\pi_j}'\hspace{-1.0mm}^{(0)},
\eeq
(with $\pi^0$, $\eta_8$ and $\eta_0$ being their own antiparticles)
as it should be.

The results above are consistent with the operators identified in
Ref.~\cite{RSWZ} following a similar reasoning. 
Except for an arbitrary phase, \eqs{eqmops} and (\ref{eqmopshf})
also agree with the result 
of diagonalizing the meson polarization function $J_{ij}$ for vanishing
quark masses (see Sect.~\ref{mesons}).
We should keep in mind, however, that we have evaluated the meson 
vertices at the pole only. In general, the diagonalization of the
T-matrix leads to 4-momentum dependent vertex functions with
4-momentum dependent weights for the contributing operators. 
For instance, the $\pi^+$ vertex can be written as
\beq
\Gamma_{\pi^+}'(q) \= -i \Big(\sin\varphi(q)\,\Gamma_{57}^{p\uparrow}
                             \;-\;\cos\varphi(q)\,\Gamma_{75}^{p\downarrow}\,
                        \Big),
\label{Gammapiq}
\eeq
with some function $\varphi(q)$ and an arbitrary overall phase which 
we have chosen to be purely imaginary to comply with \eq{eqmops}.
For equal quark masses, $\varphi(0) = \frac{\pi}{4}$ and we recover
$\Gamma_{\pi_j}'(0) = \Gamma_{\pi_j}'\hspace{-2.0mm}^{(0)}$, \eq{eqmops}.

Next, we evaluate \eq{WTING} for non-vanishing (but still equal)
quark masses $m$.
In this case the pseudoscalar vertex contributes a pole at
$\vec q = 0$ and $q_0 \rightarrow m_\pi$, which is present for any choice 
of $m_\pi$ (see \eq{polp}). Since the r.h.s. of \eq{WTING} remains
non-singular, this pole must be cancelled by the pole in the axial
part (unless $g_\pi^2\,I^{(\pi)}$ vanishes). 
This means that the formerly massless mesons receive a mass which
is determined by the requirement that the residues of the axial
and the pseudoscalar pole cancel each other. One finds
\beq
    m_\pi^2\,f_\pi^2 \= 2m\, g_\pi\,f_\pi\,I^{(\pi)}.
\label{GORex}
\eeq
This is an exact relation, valid for arbitrary values of $m$. 
We can now perform a chiral expansion of this equation to leading order.
This amounts to replace $g_\pi f_\pi$ by $\Delta$ and to evaluate 
\begin{alignat}{1}
    I^{(\pi)} = i&\intk
\nonumber \\  
    &\;\;\TrNG\left[\bar\Gamma_\pi'(q)\,  S(k+q)\,
    (i\gamma_5\,t_\pi)_\mathit{NG} S(k)\right]\Big|_{q_0=m_\pi,\vec q=0}
\end{alignat}
to leading nontrivial order in the quark mass.
It turns out that $I^{(\pi)}$ vanishes in the chiral limit and the leading 
order is linear in $m$.\footnote{If we had taken into account 
dynamical quark masses it would be linear in the constituent quark mass $M$. 
The leading term on the r.h.s. of
\eq{GOR} would then read $8AmM$.} 
Note that, in principle, $I^{(\pi)}$ depends explicitly (via the
quark propagators) and implicitly (via $m_\pi$) on $m$. 
One can show, however, that the implicit contributions vanish in
leading order. It is therefore consistent to evaluate the integral at 
$q_0=0$, and we obtain
\beq
    m_\pi^2\,f_\pi^2 \= 8\,A\,m^2 \+ \text{\it higher orders},
\label{GOR}
\eeq
with
\begin{alignat}{1}
    A = \frac{\Delta}{4}\,i
    &\intk
\nonumber \\   
     &\quad\Big(\frac{d}{dm}
    \TrNG\left[\bar\Gamma_\pi'\hspace{-1.0mm}^{(0)}\,  S(k)\,
    (i\gamma_5\,t_\pi)_\mathit{NG} S(k)\right]\Big)\Big|_{m=0}
\label{Aex}
\end{alignat}
and $\Gamma_\pi'\hspace{-1.0mm}^{(0)}\equiv\Gamma_\pi'(0)$ as given in \eq{eqmops}.
This expression can be evaluated exactly. The result is given in
\eq{eq_A_exact} in the appendix. 
Expanding that formula in $\Delta$, we find
\beq
    A \= \frac{\Delta^2}{8\pi^2}\,
    \Big( -\ln{y^2} - 2 + \frac{4}{3}\ln{2} + \ln(x^2-1) \Big)
    \+ \dots,
\label{ANJL}
\eeq
where we have introduced the abbreviations
\beq
    x = \frac{\Lambda}{\mu}, \quad y = \frac{\Delta}{\mu},
\label{xydef}
\eeq
with $\Lambda$ being the 3-momentum cutoff. 
This should be compared with the weak-coupling result~\cite{SS99},
\begin{equation}
 A_\mathit{wc}=\frac{3 \Delta^2}{4 \pi^2}.
 \label{Awc}
\end{equation}
An important difference is the logarithmic term in \eq{ANJL}, 
which does not exist in \eq{Awc}. 
In fact, in the beginning, there was a controversy about the correct 
weak-coupling limit, and similar logarithmic terms have also been discussed 
in the literature~\cite{RSWZ,Hong:1999ei,Manuel:2000wm,Beane:2000ms,BS2002,
Ruggieri:2007pi}.
To be precise, it was found that the leading contribution to $A$ in
weak coupling is given by~\cite{Beane:2000ms}
\beq
    A_{\Delta\bar{\Delta}} \= -\frac{\Delta\bar{\Delta}}{8\pi^2}\ln{y^2},
\label{ADDbar}
\eeq
where $\bar{\Delta}$ is the antiparticle gap. 
In the NJL model, the gap function is energy independent
and, hence, $\bar{\Delta}=\Delta$. Therefore, our result, \eq{ANJL},
is consistent with \eq{ADDbar}. 
Moreover, if we introduce particle and antiparticle
gaps by hand as independent constants in the quark propagator,
we recover \eq{ADDbar} in leading order.

On the other hand, as argued in the erratum of Ref.~\cite{SS99} and
confirmed in Ref.~\cite{Schafer:2001za}, in a gauge invariant treatment
of QCD at weak coupling, the antiparticle gap contributions
are cancelled by other terms, and the logarithmic terms drop
out. This finally leads to \eq{Awc}. 
Of course, gauge invariance is not an issue in the NJL model. 
Nevertheless, the logarithmic behavior at very weak coupling should
be viewed as a model artifact.

%%%%%%%%%%%%%%%%%%%%%%%%%%%%%%%%%%%%%%%%%%%%%%%%%%%%%%%%%%%%%%% 

\section{Numerical results}
\label{numerics}
In the following we present our results for $T=0$ and a fixed
quark chemical potential $\mu = 500$~MeV.

\subsection{Pion decay constant in the chiral limit}
\label{sec:fpinum}

\begin{figure}
\begin{center}
 \includegraphics[width=\linewidth]{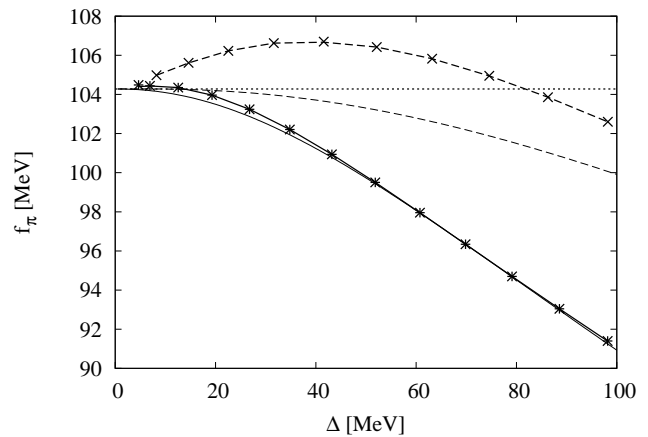}
 \caption{Pion decay constant $f_\pi$ in the chiral limit as a function
of the gap parameter $\Delta$: numerical results (points) in comparison
with the semi-analytical formula \eq{fpi2tot}, 
neglecting (thin lines) and including (thick lines) 
the momentum dependence of the vertex function.
Asterisks and solid lines: $\Lambda=600$\,MeV;
crosses and dashed lines: $\Lambda=700$\,MeV;
dotted line: weak-coupling limit, \eq{fpiwc}.}
\label{fig:fpi}
\end{center}
\end{figure}

We begin with a discussion of the pion decay constant in the 
chiral limit, i.e., for vanishing quark masses. 
In \fig{fig:fpi} our results are displayed as functions of the gap 
parameter $\Delta$ for two different values of the cutoff $\Lambda$.
The points correspond to the numerical results for $\Lambda = 600$~MeV 
(asterisks) and $\Lambda = 700$~MeV (crosses).
To be precise, these calculations have been performed
with $m_u = m_d = m_s = 0.1$~MeV for technical reasons.

%%%%%%%%%%%%%%%%%%%%%%%%%%%%%%%%%%%%%%%%%%%%%%%%%%%%%%%%%%%%%%%%%%%%%%%%
%%%%%%%%%%%%%%%%%%%%%%%%%%%%%%%%%%%%%%%%%%%%%%%%%%%%%%%%%%%%%%%%%%%%%%%%%%

The weak-coupling limit, $\Delta \rightarrow 0$, of $f_\pi$ has been
derived in Ref.~\cite{SS99} from an effective theory involving 
only fermionic modes in the vicinity of the Fermi surface. 
The result, 
\begin{equation} 
f_\pi^2 \;\rightarrow\; \frac{21 - 8\ln 2}{18} \frac{\mu^2}{2 \pi^2},
\label{fpiwc}
\end{equation}
is universal and should hold in any model exhibiting the same symmetry
pattern.\footnote{Strictly speaking, the chemical potential $\mu$
should be replaced by the Fermi momentum $p_F$ because in general the
Fermi velocity can differ from the speed of light.}  Indeed, our
results converge to this limit for $\Delta \rightarrow 0$.
In the general case we find, however, deviations from \eq{fpiwc}, e.g.,
about 10\% for $\Lambda = 600$~MeV and $\Delta \approx
80$~MeV. Moreover, these deviations depend rather strongly on the
cutoff.

In order to understand this behavior and to confirm the correct 
weak-coupling limit, we employ the semi-analytical formula
derived in App.~\ref{appfpi},
\beq
    f_\pi^2 \= \tilde f_\pi^2 \+ \delta f_\pi^2.
\label{fpi2tot}
\eeq
$\tilde f_\pi^2$ thereby describes the contribution to $f_\pi$ arising
for a constant pion vertex function, i.e., neglecting the energy
dependence of the mixing angle $\varphi$ in \eq{Gammapiq}.
This part is given in a closed analytical form in \eq{fjeq}.
$\delta f_\pi^2$, on the other hand, incorporates the effect of the 
energy dependence of the vertex function and is given in \eq{fpimomentum}.
It is proportional to the derivative $d\varphi/dq_0$, 
which is evaluated numerically.

Expanding $\tilde f_\pi^2$ and the analytical factor of 
$\delta f_\pi^2$ for small values of $\Delta$ yields
\beq
\tilde f_\pi^2 =  \frac{\mu^2}{36 \pi^2} 
\Big\{ (21 - 8\ln{2}) \;-\;9y^2\ln{y^2} \;-\; c_2 y^2 \+ \dots \Big\}
\label{fpi0exp}
\eeq
and 
\beq
    \delta f_\pi^2 \= \frac{\mu^2}{36\pi^2}
    \Big\{-36y^2\ln y^2 - d_2 y^2 + \dots \Big\} 
    \,\mu\frac{d\varphi}{dq_0}\Big|_{q=0},
\label{deltafpiexp}
\eeq
where
\begin{alignat}{1}
    c_2 \=& \frac{81}{4} - 18\ln{2} - 9\ln{(x^2-1)} + 
    \frac{45x^2-27}{(x^2-1)^2},
\nonumber\\
    d_2 \=& 102 - 56\ln 2 - 36\ln(x^2-1) + \frac{36}{x^2-1},
\end{alignat}
and $x$ and $y$ are defined in \eq{xydef}. 

We see that the weak-coupling limit, \eq{fpiwc}, is correctly reproduced
by the leading term in $\tilde f_\pi^2$, while $\delta f_\pi^2$ does not
contribute to this order, provided $d\varphi/dq_0$ does not diverge as
strongly as $(y^2\ln y^2)^{-1}$ for $\Delta \rightarrow 0$.
We therefore expect that $\tilde f_\pi^2$ gives the main contribution
to $f_\pi$.

As already mentioned, the weak-coupling limit is universal and
therefore must be cutoff independent.  In the detailed calculations,
this results from the fact that in the limit $\Delta\to 0$ the
integrand of the 3-momentum integral in \eq{fpiintappa} becomes
proportional to $\delta$-functions at the Fermi surface.  While the
$y^2\ln{y^2}$-term in \eq{fpi0exp} is cutoff independent as well, the
quadratic term is not universal and dependends on $\Lambda$ via the
variable $x$. This is also the case for the quadratic term in
\eq{deltafpiexp}. Anyway, the situation for $\delta f_\pi^2$ is more
complicated because the derivative $d\varphi/dq_0$ depends on the cutoff
as well.

The results of the semi-analytical formula for $f_\pi$ are indicated
by the thick lines in \fig{fig:fpi}. The solid line corresponds to 
$\Lambda = 600$~MeV and the dashed line to $\Lambda=700$~MeV.
We have employed the exact formulas, \eqs{fjeq} and (\ref{fpimomentum}),
for  $\tilde f_\pi^2$ and $\delta f_\pi^2$, respectively, 
with $d\varphi/dq_0$ being computed numerically. Obviously, the results 
for $f_\pi$ are in perfect agreement with the numerical computations.

In order to analyze the influence of the momentum dependence of the
vertex function, we display the function $\tilde f_\pi$, too (thin lines).
Since $\tilde f_\pi$ contains the leading term, \eq{fpiwc},
it is not surprising that it is the dominant contribution to $f_\pi$.
However, for a correct description of the {\it deviations} from the 
weak-coupling limit, $\delta f_\pi^2$ can be quite important:
Whereas for $\Lambda = 600$~MeV, we find that $f_\pi$
is well reproduced by $\tilde{f}_\pi$, this is not the case for
$\Lambda = 700$~MeV.  This indicates that the correction due to
the contributions arising from the momentum dependence
of the vertex function are rather small for $\Lambda = 600$~MeV and
considerably larger for $\Lambda = 700$~MeV.

%%%%%%%%%%%%%%%%%%%%%%%%%%%%%%%%%%%%%%%%%%%%%%%%%%%%%%%%%%%%%%%%%%%
\begin{figure}
\begin{center}
 \includegraphics[width=\linewidth]{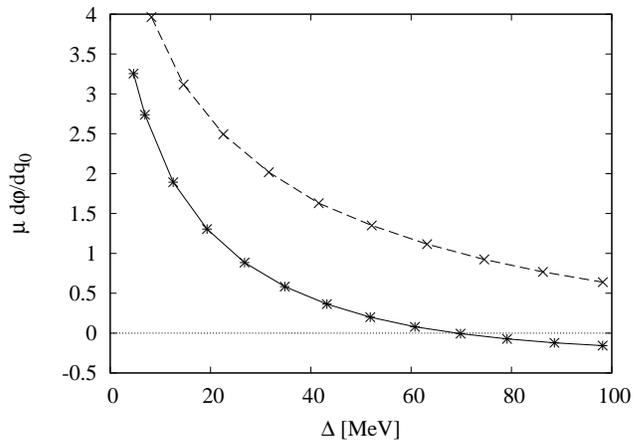}
 \caption{Derivative of the mixing angle $\varphi$ as a function of 
 the gap parameter $\Delta$.
 Solid: $\Lambda = 600$~MeV, dashed: $\Lambda = 700$~MeV.}
\label{fig:mixang}
\end{center}
\end{figure}
%%%%%%%%%%%%%%%%%%%%%%%%%%%%%%%%%%%%%%%%%%%%%%%%%%%%%%%%%%%%%%%%%%%%

This result becomes clear if we look at the derivative
$d\varphi/dq_0$ at $q=0$, which is shown in
\fig{fig:mixang}.  
For $\Lambda=700$~MeV (dashed line) the derivative is nowhere small
in the shown region. Therefore, $\delta f_\pi^2$ can never
be neglected, in agreement with our findings in \fig{fig:fpi}.
On the other hand, for $\Lambda=600$~MeV (solid line) we find that
the derivative is rather small for $\Delta \gtrsim 40$~MeV,
explaining why $\delta f_\pi^2$ is negligible in this regime.  For
smaller values of $\Delta$ the correction becomes larger, in agreement with 
the deviations we found in \fig{fig:fpi}. In fact, for $\Delta \rightarrow
0$, the derivative even seems to diverge. If it was stronger than
$(y^2\ln y^2)^{-1}$, the divergence could affect the weak-coupling
limit. However, at least numerically, we find that $d\varphi/dq_0$ grows 
much slower than $(y^2\ln y^2)^{-1}$, and we therefore conclude that the
weak-coupling limit is safe.

%%%%%%%%%%%%%%%%%%%%%%%%%%%%%%%%%%%%%%%%%%%%%%%%%%%%%%%%%%%%%%%%%
\subsection{Equal quark masses}
\label{sec:eqmassnum}
%%%%%%%%%%%%%%%%%%%%%%%%%%%%%%%%%%%%%%%%%%%%%%%%%%%%%%%%%%%%%%%%%

Next, we study the effect of explicit chiral symmetry breaking 
through non-vanishing, but equal quark masses $m_u = m_d = m_s \equiv m$. 
For the cutoff, we choose $\Lambda = 600$~MeV and keep this value fixed
in the remainder of this article.

\begin{figure}
\begin{center}
 \includegraphics[width=\linewidth]{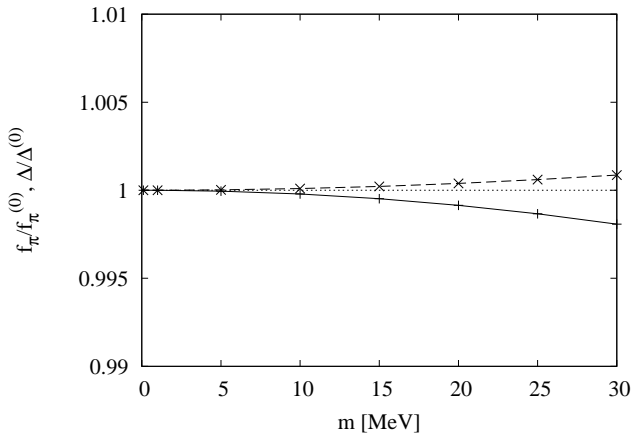}
 \caption{$f_{\pi}$ (solid line) and $\Delta$ (dashed line)
 divided by their respective chiral limit values 
 $f_{\pi}^{(0)}$ and $\Delta^{(0)}$
 as functions of the common quark mass $m$.
 The calculations have been performed for $H \Lambda^2 =1.4$,
 corresponding to $\Delta^{(0)} = 79.1$~MeV.
 \label{fig:fpi_mass}
 }
\end{center}
\end{figure}

Before turning to our main focus, i.e., the masses of the Goldstone 
bosons, we briefly investigate the dependence of the pion decay constant 
and of the gap parameter $\Delta$ on $m$. The results are displayed in 
\fig{fig:fpi_mass}.
We find that both quantities depend only very weakly on the quark mass.
In the plotted range they vary less than 0.2\,\%.
This is much weaker than the $m$ dependence of $f_\pi$ in vacuum
or of the constituent quark mass in comparable models.  
In the following discussion we will therefore neglect the distinction
between the $m$ dependent gap parameter $\Delta$ and its chiral limit
$\Delta^{(0)}$ and often use $\Delta$ in order to characterize the
coupling strength.

%%%%%%%%%%%%%%%%%%%%%%%%%%%%%%%%%%%%%%%%%%%%%%%%%%%%%%%%%%%%%%%%%%%%
%%%%%%%%%%%%%%%%%%%%%%%%%%%%%%%%%%%%%%%%%%%%%%%%%%%%%%%%%%%%%%%%%%%%

\begin{figure}
\begin{center}
 \includegraphics[width=\linewidth]{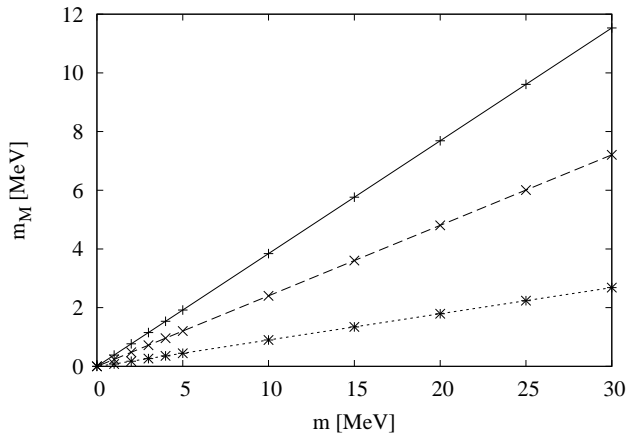}
 \caption{Masses of the flavored (pseudo-) Goldstone bosons
as functions of a common quark mass $m$ for three different
couplings. Solid: $H\Lambda^2 = 1.4$ (corresponding to 
$\Delta = 79.1$~MeV), 
dashed: $H\Lambda^2 = 1.0$ ($\Delta = 43.2$~MeV), 
dotted: $H\Lambda^2 = 0.6$ ($\Delta = 12.5$~MeV). 
\label{fig:mmeq}
}
\end{center}
\end{figure}

From \eq{GOR}, we expect that, to leading order, 
the masses $m_M$ of the Goldstone bosons in the octet depend linearly 
on $m$,
\beq
    m_{M} \= \sqrt{\frac{8A}{f_\pi^2}}\;m~.
    \label{eq:m_M}
\eeq
This is confirmed by our numerical calculations. 
In \fig{fig:mmeq} the values of $m_{M}$ are displayed as functions 
of the quark mass for three different couplings $H\Lambda^2 = 0.6$, 
$1.0$, and $1.4$, corresponding to CFL gaps of 12.5~MeV, 43.2~MeV, 
and 79.1~MeV, respectively.
As one can see, the results show an almost perfect linear behavior.
%%%%%%%%%%%%%%%%%%%%%%%%%%%%%%%%%%%%%%%%%%%%%%%%%%%%%%%%%%%%%%%%%%%
%%%%%%%%%%%%%%%%%%%%%%%%%%%%%%%%%%%%%%%%%%%%%%%%%%%%%%%%%%%%%%%%%%%%

The slopes of the straight lines are decreasing 
with decreasing coupling strength $H$, i.e., with decreasing $\Delta$.
This is also expected from the QCD weak-coupling limit,
\eq{Awc} and \eq{fpiwc} inserted in \eq{eq:m_M}.
However, as discussed in Sect.~\ref{sect:CWTI}, we expect 
that the $\Delta$-dependence of the slopes in the NJL model and in 
weak-coupling QCD is rather different.
To study this aspect in detail, we determine the slopes $a$ of the
functions $m_M(m) = a\,m$ for different values of $\Delta$ 
and use our ($\Delta$ dependent) results for $f_\pi$ to obtain
$A = \frac{1}{8}a^2f_\pi^2$. 

\begin{figure}
\begin{center}
 \includegraphics[width=\linewidth]{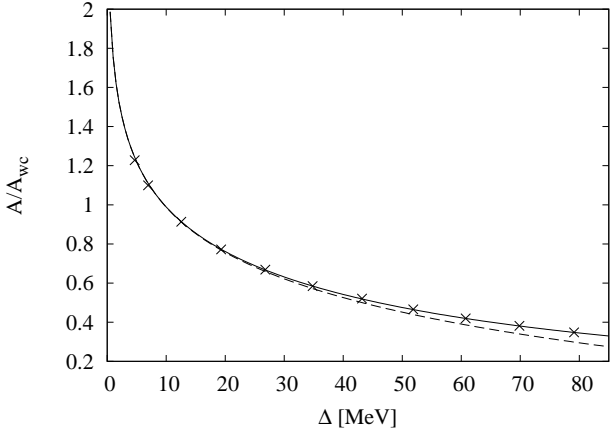}
 \caption{Ratio of $A$ and $A_\mathit{wc}$ as a function of the gap 
parameter $\Delta$. Points: numerical results; solid line: 
exact analytical result with $A$ from \eq{eq_A_exact};
dashed line: approximate result with $A$ from \eq{ANJL}.}
\label{fig:A_Delta}
\end{center}
\end{figure}

The result, divided by the weak-coupling limit $A_\mathit{wc}$ is
displayed in \fig{fig:A_Delta} as a function of $\Delta$. 
The ratios obtained from the fit to the numerically determined meson 
masses are indicated by the crosses. 
The solid line corresponds to \eq{eq_A_exact}, i.e., to 
the exact analytical solution of 
\eq{Aex}, the dashed line indicates the approximate formula \eq{ANJL}.
The former is again in perfect agreement with the numerical results. 

We see that for small values of $\Delta$, $A$ is larger than $A_\mathit{wc}$.
This is due to the logarithmic term discussed
in Sec.~\ref{sect:CWTI} (see \eq{ANJL} and the subsequent discussion). 
For large couplings, on the other hand, the NJL-model value of $A$ is 
considerably smaller than $A_\mathit{wc}$, leading to 
even smaller Goldstone-boson masses than predicted in weak coupling.
This has also been found in
Ref.~\cite{Buballa:2004sx} within a rather different approach.

%%%%%%%%%%%%%%%%%%%%%%%%%%%%%%%%%%%%%%%%%%%%%%%%%%%%%%%%%%%%%%%%%%%
%%%%%%%%%%%%%%%%%%%%%%%%%%%%%%%%%%%%%%%%%%%%%%%%%%%%%%%%%%%%%%%%%%%

\subsection{Unequal quark masses}

\begin{figure}
\begin{center}
 \includegraphics[width=\linewidth]{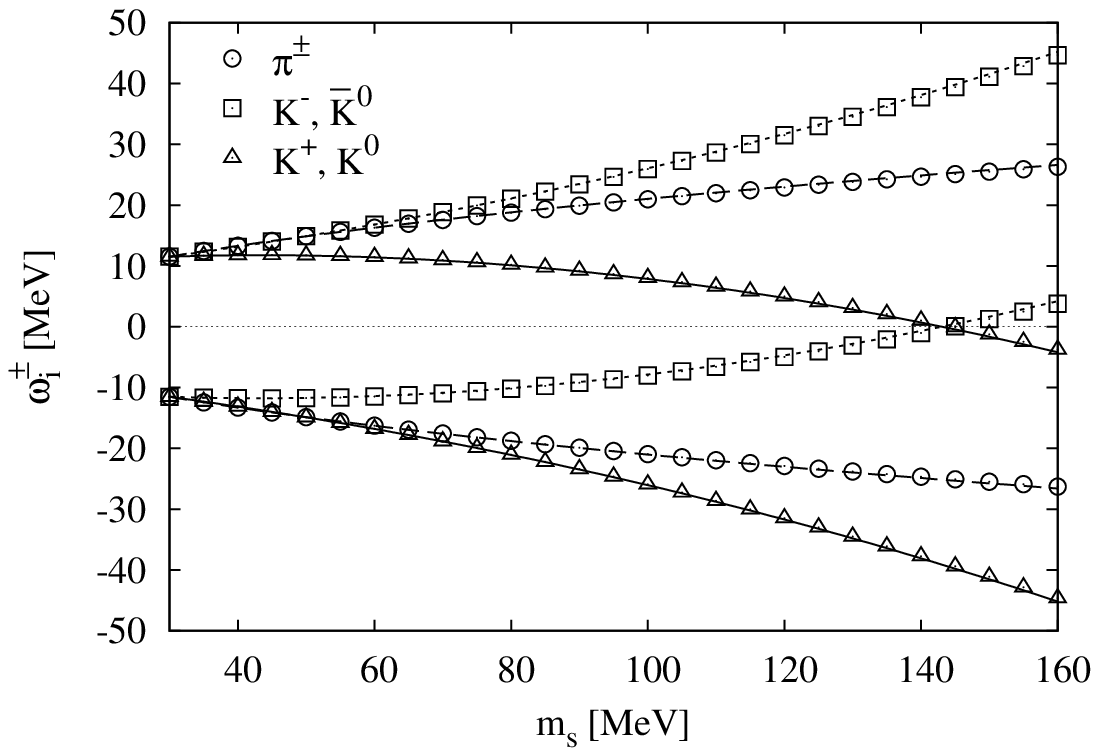}
 \includegraphics[width=\linewidth]{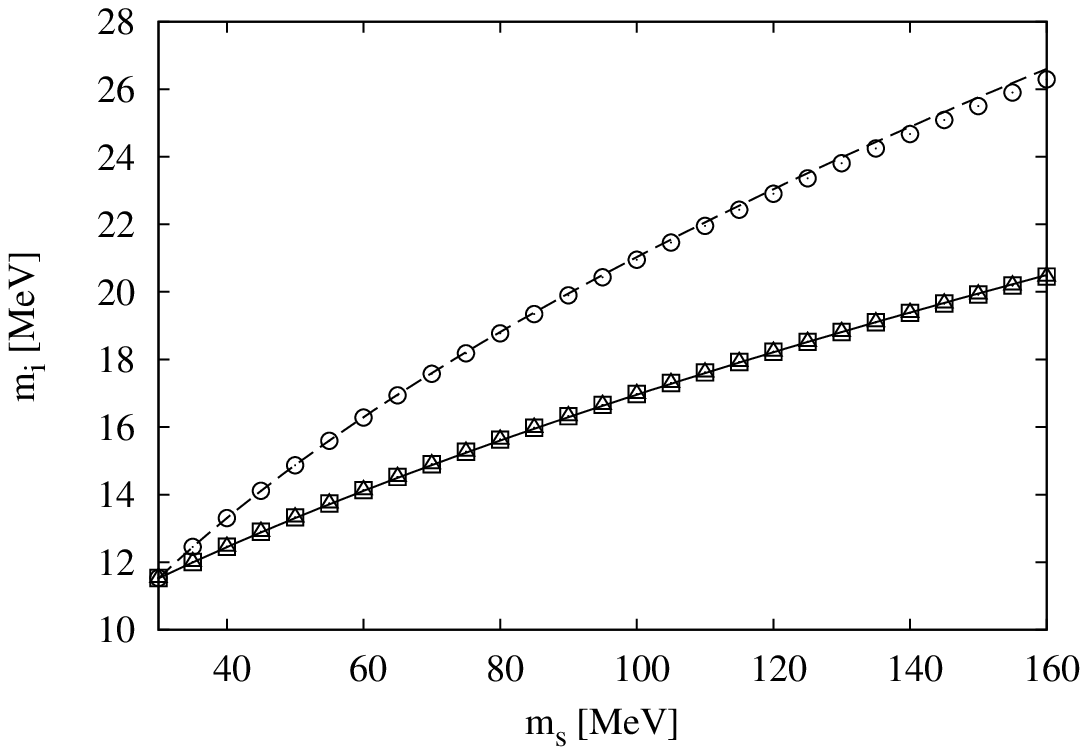}
 \includegraphics[width=\linewidth]{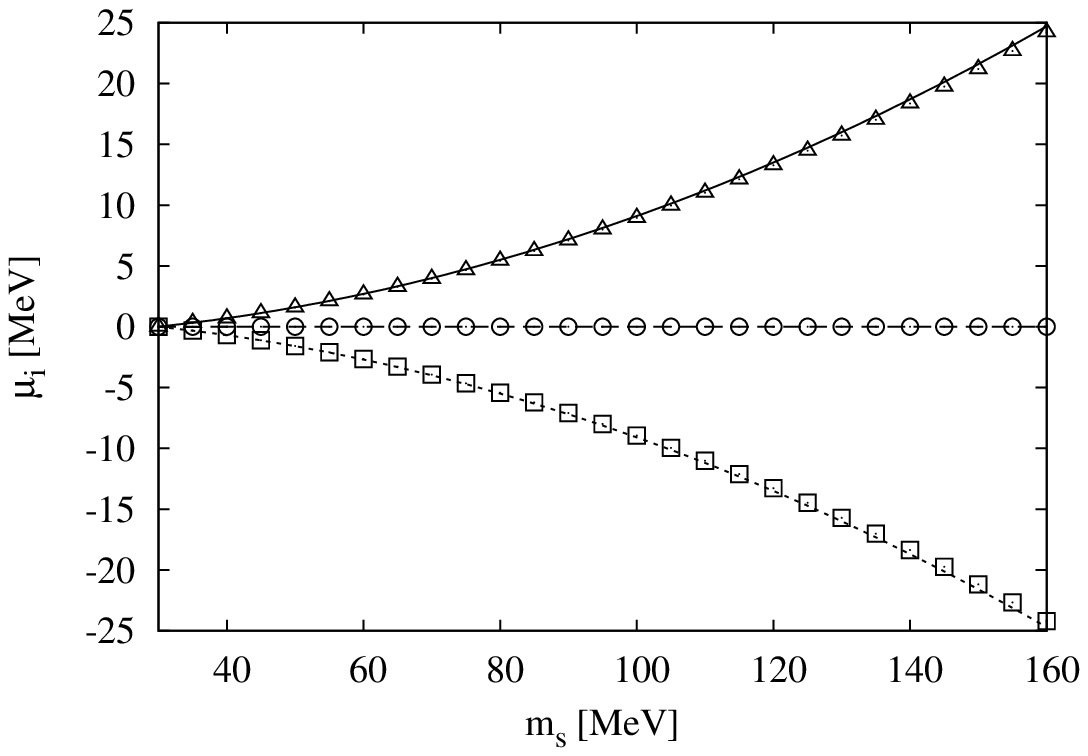}
 \caption{Properties of the flavored Goldstone modes 
  as functions of the strange quark mass for $m_u = m_d = 30$~MeV
  and $H\Lambda^2 = 1.4$ ($\Delta = 79.1$~MeV): 
  pole positions of the T-matrix at $\vec q = 0$ (upper panel),
  meson masses (center), and meson chemical potentials (lower panel). 
  The various points indicate the numerical results:
  $\pi^\pm$ (circles),  $K^+$ and $K^0$ (triangles), and
  $K^-$ and $\bar{K}^0$ (squares). The lines correspond to the
  predictions from \eqs{omegaipm}, (\ref{mieft}), and (\ref{muieft}).
\label{fig:mmdif}
}
\end{center}
\end{figure}

Finally, we study the effect of unequal quark masses.
In the upper panel of \fig{fig:mmdif} the pole positions of the flavored 
(pseudo-) Goldstone modes are displayed as functions of the strange quark 
mass, keeping $m_u$ and $m_d$ fixed. Our numerical results are indicated
by the points. For practical reasons,
we have chosen a relatively strong diquark coupling $H\Lambda^2 = 1.4$ 
(corresponding to $\Delta = 79.1$~MeV) and a relatively large value
of 30~MeV for $m_u$ and $m_d$, in order to have not too small meson 
masses.
As one can see, the poles which are degenerate for equal masses
split into three branches,
corresponding to different strangeness, i.e., pions ($S=0$), 
kaons ($S=+1$), and antikaons ($S=-1$).
On the other hand, since we have chosen $m_u = m_d$, the different
isospin states of these modes, i.e., $\pi^-$ and $\pi^+$, 
$K^+$ and $K^0$, and $K^-$ and $\bar{K}^0$ remain degenerate.

%%%%%%%%%%%%%%%%%%%%%%%%%%%%%%%%%%%%%%%%%
%%%%%%%%%%%%%%%%%%%%%%%%%%%%%%%%%%%%%%%%%%%%%%%%%%

Our results can be analyzed in terms of the parameterization given 
in \eq{TMgen}. 
Each mode  $T^{(i)}$ has two poles, which for $\vec q = 0$ are
located at
\beq
    q_{0} \= \pm m_i \,-\, \mu_i \;\equiv\;
    \omega_i^\pm.
\label{omegaipm}
\eeq
We can thus extract the meson masses and chemical potentials as
\beq
    m_i \= \frac{1}{2}(\omega_i^+ - \omega_i^-), \qquad    
   \mu_i \= -\frac{1}{2}(\omega_i^+ + \omega_i^-).
\eeq
The resulting functions are displayed in the two lower panels of
\fig{fig:mmdif}. The masses are moderately rising with $m_s$ and
exhibit an ``inverse ordering'' ($m_K < m_\pi$), as predicted first
in Ref.~\cite{SS99}. Also note that kaons and antikaons have the 
same masses.
For the meson chemical potentials, on the other hand, we find
$\mu_K = -\mu_{\bar K}$, as it should be, and $\mu_\pi = 0$.

These results can be compared with those derived in Refs.~\cite{SS99}
and \cite{BS2002,Kaplan:2001qk} in an effective field theory (EFT) 
approach. For $m_u = m_d \equiv m_q$, they read
\beq
    m_{\pi} = \sqrt{\frac{8A}{f_{\pi}^2} m_s m_q}~, \quad
    m_{K} = m_{\bar{K}} = \sqrt{\frac{4A}{f_{\pi}^2} m_q (m_q + m_s)},
\label{mieft}
\eeq
and
\beq
    \mu_{\pi^\pm} \= 0~, \qquad 
    \mu_{K^\pm}  \= \mu_{K^0,\bar{K}^0} \= \pm \frac{m_s^2 - m_q^2}{2 \mu}.
\label{muieft}
\eeq
For $A$ and $f_\pi$ we insert the NJL-model values obtained in 
Sects. \ref{sec:fpinum} and \ref {sec:eqmassnum} in the limit of
vanishing quark masses.
The resulting functions are indicated by the lines in \fig{fig:mmdif}.
Obviously, they agree almost perfectly with the numerical calculations. 
Note, however, that the agreement would not be good if we had used
the weak-coupling results for $A$ and $f_\pi$.

When $\mu_i$ reaches the value of $m_i$,
\beq
    \mu_i(m_q,m_s^\mathit{crit}) \= m_i(m_q,m_s^\mathit{crit})~,
\label{mscritdef}
\eeq
$\omega_i^+$ vanishes and meson condensation sets in. For the chosen 
parameters, this occurs in the kaon branch ($K^+$ and $K^0$) at 
a critical strange quark mass $m_s^\mathit{crit}\approx 145\,\text{MeV}$.
For higher strange quark masses the CFL phase is no longer the correct ground 
state and the shown results have no physical meaning. 

Inserting the EFT expressions \eqs{mieft} and  (\ref{muieft}) into
\eq{mscritdef} one finds that the critical strange quark mass for 
kaon condensation is approximately given by \cite{BS2002}
\beq
    m_s^\mathit{crit} \;\approx\; 
    \left(\frac{16\,\mu^2 A}{f_\pi^2}\right)^{1/3}\,m_q^{1/3},
\label{mscritapp}
\eeq
which becomes exact in the limit $m_q \rightarrow 0$.
For $m_q = 0$ this implies that kaon condensation is favored for
arbitrarily small strange quark masses.
One might expect that this is also the case in
our model, after we found good agreement with the EFT predictions in 
\fig{fig:mmdif}.

On the other hand, this would contradict an earlier NJL-model study
\cite{Buballa:2004sx}, where a nonzero critical strange quark 
mass was found, even for $m_q = 0$.
This was concluded, without explicit construction of the Goldstone 
bosons, by studying the stability of the CFL ground state against 
partially rotating the scalar diquark condensates into pseudoscalar 
ones. Thus, with our present approach, we can check this result from 
a different perspective.
Since in Ref.~\cite{Buballa:2004sx} indications were found that the
deviations from the EFT predictions are due to terms of higher order
in the interaction, we perform our analysis using a rather strong
diquark coupling, $H\Lambda^2 = 1.7$, corresponding to 
$\Delta = 107.9$~MeV.

%%%%%%%%%%%%%%%%%%%%%%%%%%%%%%%%%%%%%%%%%
\begin{figure}
\begin{center}
 \includegraphics[width=\linewidth]{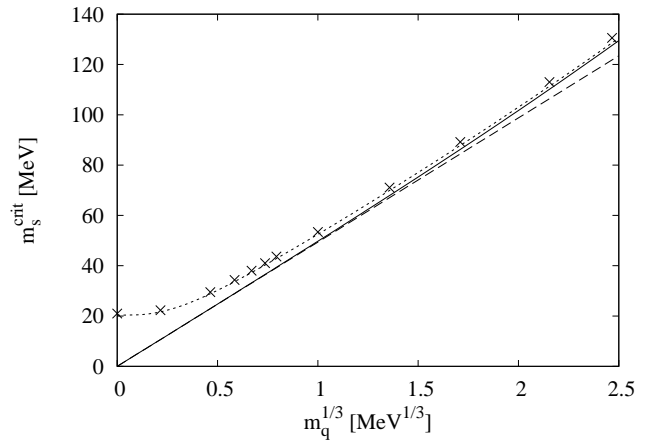}
 \caption{Critical strange quark mass $m_s^\mathit{crit}$ for kaon condensation
  as a function of $m_q^{1/3}$, where $m_q \equiv m_u =m_d$:
  NJL-model results (crosses), EFT predictions using \eqs{mieft} and
  (\ref{muieft}) (solid line), and approximate solution \eq{mscritapp}.
  The dotted line is based on \eq{mieftext} with $a_s = 0.0203$. 
  The NJL-model calculations have been performed with $H\Lambda^2 = 1.7$,
  corresponding to $\Delta = 107.9$~MeV.}
 \label{fig:kaon_cond}
\end{center}
\end{figure}
%%%%%%%%%%%%%%%%%%%%%%%%%%%%%%%%%%%%%%%%%

In \fig{fig:kaon_cond}, the critical strange quark mass for kaon condensation
is displayed as a function of the third root of $m_q$. The NJL-model
results are indicated by the crosses. We also show the solution of
\eq{mscritdef} for the EFT masses and chemical potentials \eqs{mieft} 
and (\ref{muieft}) (solid line) and the approximate solution \eq{mscritapp}
(dashed). 

We find that the NJL points are always above the EFT predictions
(solid line). However, while the deviations are small for 
$m_q \gtrsim 1$~MeV, they become essential for smaller values of $m_q$.
In particular, we confirm that $m_s^\mathit{crit}$ does not vanish at
$m_q = 0$ but goes to a finite value, which is about 21~MeV in our 
example.

%%%%%%%%%%%%%%%%%%%%%%%%%%%%%%%%%%%%%%%%%
\begin{figure}
\begin{center}
 \includegraphics[width=\linewidth]{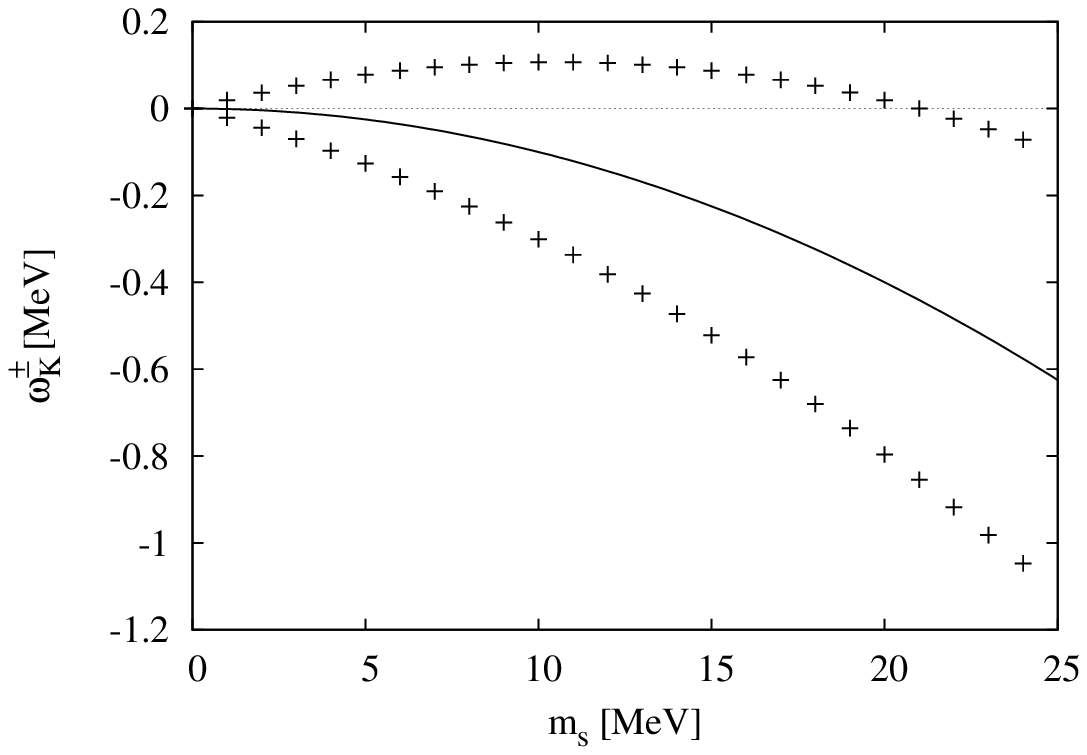}
 \includegraphics[width=\linewidth]{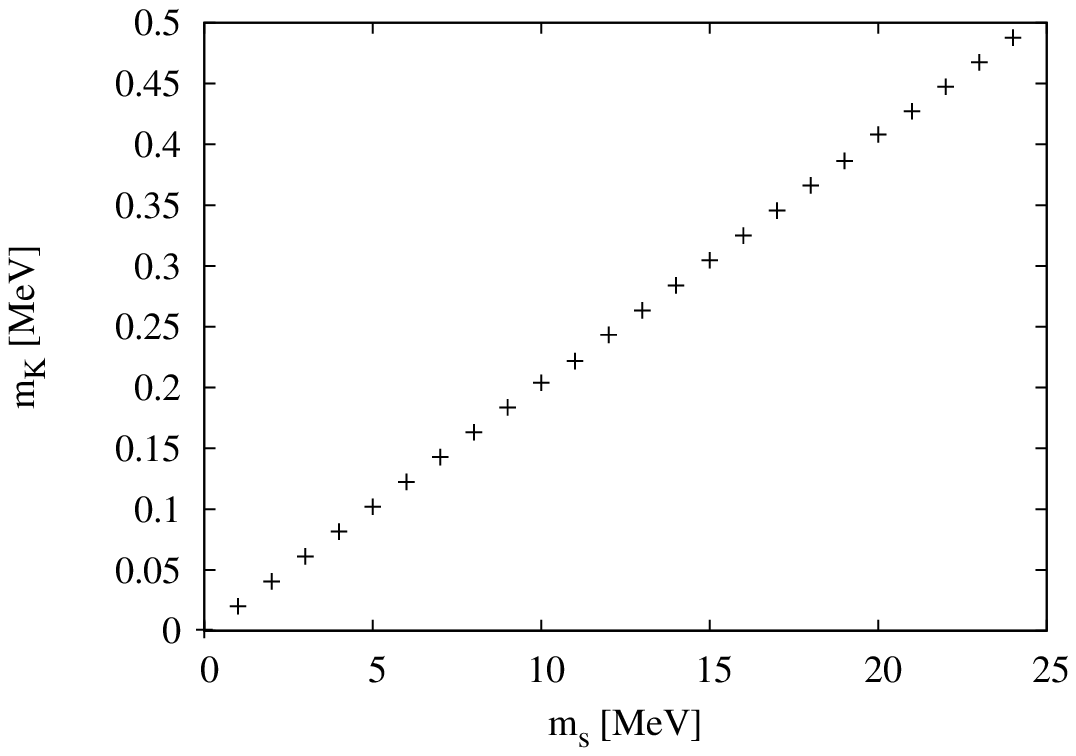}
 \includegraphics[width=\linewidth]{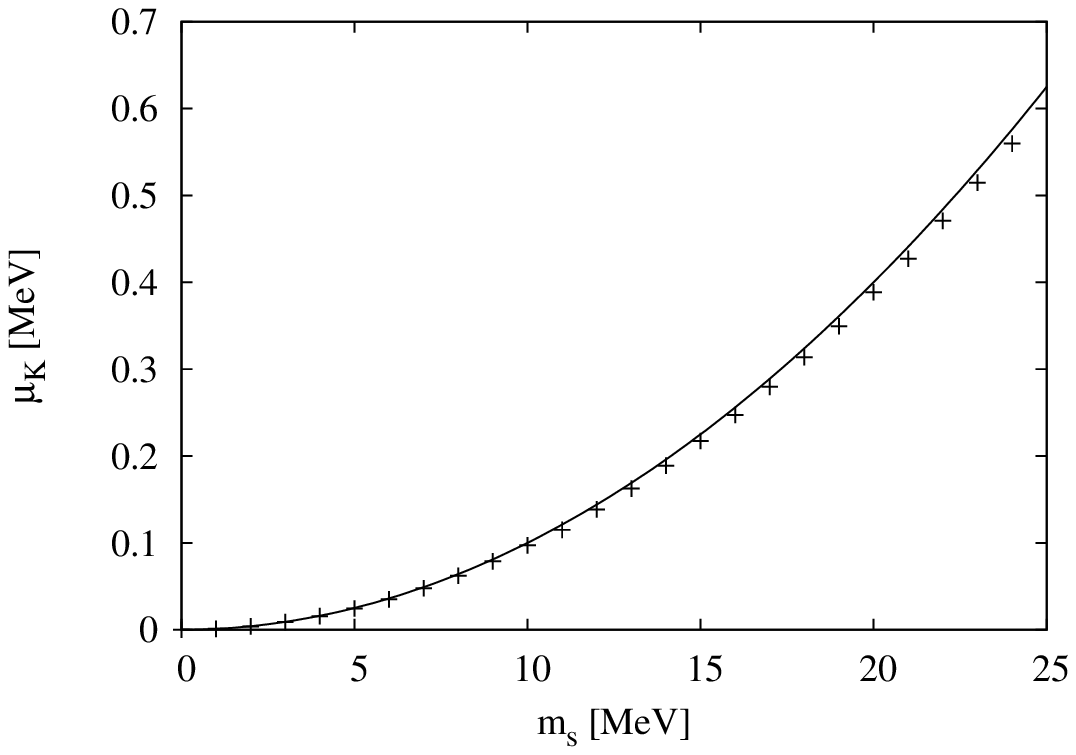}
 \caption{Kaon ($K^+$ and $K^0$) properties as functions of the strange 
  quark mass for $m_u = m_d = 0$ and $H\Lambda^2 = 1.7$ ($\Delta = 107.9$~MeV): 
  pole positions of the T-matrix at $\vec q = 0$ (upper panel),
  kaon masses (center), and kaon chemical potentials (lower panel). 
  The crosses indicate the numerical results, the lines correspond to 
  the predictions from \eqs{omegaipm}, (\ref{mieft}), and (\ref{muieft}).
  Note that from \eq{mieft} $m_K$ is expected to vanish.
\label{fig:K_pl_H1.7} 
}
\end{center}
\end{figure}
%%%%%%%%%%%%%%%%%%%%%%%%%%%%%%%%%%%%%%%%%%%%%%%%%%

To understand this behavior, we analyze the poles of the T-matrix
in the kaon channel for $m_q = 0$ as functions of $m_s$.
In \fig{fig:K_pl_H1.7} the NJL-model results (crosses) are compared 
with the EFT predictions (solid lines). 
As before, the pole positions (upper panel) can be interpreted in terms 
of  kaon masses (center) and chemical potentials (lower panel). 
It turns out that the latter are in fair agreement with \eq{muieft}.
On the other hand, while \eq{mieft} predicts the kaon masses to be
zero for $m_q = 0$, we find that $m_K$ is in general non-zero and
rises linearly with $m_s$. 
As a consequence, the pole position $\omega_K^+$ is not degenerate
with $\omega_K^-$ (cf. \eq{omegaipm}) and first rises with $m_s$.
Hence, kaon condensation does not occur
at arbitrarily small values of $m_s$ but only for $m_s \gtrsim 21$~MeV, as
already seen in \fig{fig:kaon_cond}.

It should be noted, however, that, although the linear rise of $m_K$ 
with $m_s$ is qualitatively different from \eq{mieft}, the slope is 
very small on a quantitative scale. 
In our example, a linear fit $m_K = a_s m_s$ yields 
$a_s = 0.0203$, which is an order of magnitude smaller than
the slope in \eq{mieft}, $\sqrt{4A/f_\pi^2} = 0.347$.
Moreover, by varying the coupling strength, we found numerically
that $a_s$ depends quadratically on $\Delta$.
This is consistent with our expectation that the effect 
corresponds to a higher-order correction in the interaction
and only becomes visible when the leading order, \eq{mieft}, 
is artificially suppressed by choosing very small values of $m_q$. 
In fact, for any realistic value of $m_u$ and $m_d$ the correction
is quite irrelevant. 

Also note that our results are somewhat complementary to 
Ref.~\cite{Ruggieri:2007pi}, where corrections to the effective
kaon chemical potential have been discussed. However, it was found 
there that these corrections are of the order of 
$(m_s^2/2\mu\Delta)^2$, which is completely negligible in all of 
our examples. 

We may finally combine \eq{mieft} with our numerical findings for $m_q=0$
by parameterizing
\beq
    m_{K} = \sqrt{\frac{4A}{f_{\pi}^2} m_q (m_q + m_s) + a_s^2 m_s^2}.
\label{mieftext}
\eeq
Equating this with the kaon chemical potential, \eq{muieft}, we obtain
the dotted line in \fig{fig:kaon_cond} for the critical strange quark 
mass. It is obviously in good agreement with our numerical results.
The tiny deviations for small values of $m_q$ can be explained by the
fact that the kaon chemical potential in \fig{fig:K_pl_H1.7} is slightly
overestimated by \eq{muieft}. A similar effect could also play a role
at larger $m_q$, but there might be other terms as well.

%%%%%%%%%%%%%%%%%%%%%%%%%%%%%%%%%%%%%%%%%%%%%%%%%%%
\section{Summary and conclusions}
\label{summary}

We have studied the properties of pseudoscalar Goldstone boson
excitations in the color-flavor-locked phase within an NJL-type
model. To that end we solved the Bethe-Salpeter equation in RPA.
So far we have only included quark-quark interactions such that our
Goldstone boson states are in fact superpositions of
diquark and di-hole states. 

Our results are consistent with the model independent predictions
of the low-energy effective theory, i.e., with those predictions
which only depend on the symmetry breaking pattern.
We found, however, deviations from the values for the constants
appearing in the LEET, as for instance the pion decay constant
$f_\pi$, obtained in the weak-coupling limit. 
In fact, it was the main motivation of this paper to locate such 
deviations and to understand their origins. In several cases
this could even be
done analytically. Although these model results are in general 
not universal, they may give important hints about to what extent
the weak-coupling results can be trusted in the intermediate-density
regime and where to expect major deviations.   

The weak-coupling limit for $f_\pi$~\cite{SS99} is correctly
reproduced in the chiral limit in zeroth order in the gap parameter
$\Delta$. This must be the case since this result is universal,
i.e., independent of the specific choice of the interaction.  For
$\Delta$ not being small we found deviations, typically of the order of
a few percent. We have shown that this is an effect of higher order in
the gap parameter.  In this context, the momentum dependence of the
dressed vertex function plays an interesting role.

Next, we discussed the masses of the Goldstone bosons in the 
limit of equal quark masses. In agreement with the LEET 
prediction, we found that the meson masses 
behave linearly in the quark masses. However, the corresponding coefficient
$A$ does not agree with the weak-coupling result obtained in HDET \cite{SS99}.
In the limit $\Delta\to 0$, this should be viewed as an artifact of the NJL 
model. Probably more relevant are
the deviations at large values of $\Delta$ where we found the meson masses
to be considerably smaller than predicted by the weak-coupling formula.

Finally, we have studied the case of unequal quark masses. 
In general, we found a very good agreement
with the LEET prediction for the meson masses
and effective chemical potentials, \eqs{mieft} and (\ref{muieft}).
In particular, we
found $m_K < m_\pi$, i.e., an inverse meson mass ordering as predicted 
in Ref.~\cite{SS99}. We also confirmed that the strange quark mass acts as
an effective strangeness chemical potential, eventually leading
to kaon condensation at sufficiently large values of $m_s$ 
Refs.~\cite{BS2002,Kaplan:2001qk}.
However, even quantitatively, our model results are in an almost
perfect agreement with \eqs{mieft} and (\ref{muieft}) if the constants
$A$ and $f_\pi$ entering \eq{mieft} are taken from our preceding 
NJL-model studies in the limit of vanishing quark masses. 
Since these values do not agree with the weak-coupling limit, 
as discussed above, our meson masses are in general smaller than those 
obtained with the weak-coupling coefficients. 

In the limit of vanishing light quark masses, we found a qualitative 
difference. The critical strange quark mass
for the onset of kaon condensation does not vanish with the cubic root
of light quark masses but attains a nonzero value. We identified this
numerically as a higher order effect on the kaon mass,
adding a $\Delta^2 m_s$ dependence at low light quark masses. 

This paper should be seen as the basis for further studies of the
Goldstone boson dynamics in cold dense quark matter at non-asymptotic
densities where deviations from the weak-coupling limit become
visible. As already mentioned,
our simple Lagrangian does not include
quark-antiquark interactions, which, although subdominant, can give
important corrections to the pseudoscalar meson masses. Moreover, we 
miss instanton effects, which have been shown to be important
\cite{Manuel:2000wm,Yamamoto:2007ah}. 
The ultimate goal is to include the back reaction
of the Goldstone bosons on the phase structure of color superconducting 
quark matter \cite{Nickel:2007}. 
In the intermediate density regime, the ratio of the gap 
parameter and the Fermi energy is of the order of 0.25, such that the 
Goldstone boson excitations can have a significant effect on the ground 
state properties. 

\begin{acknowledgments}

We thank M. Ruggieri for useful comments.
This work has been supported in part by the BMBF under contract 06DA123
and by the Helmholtz-University Young Investigator Grant VH-NG-332.
 
\end{acknowledgments}

\appendix

\section{Gap equations}
\label{appgap}

The Dyson equation shown in Fig.~\ref{fig:dyson_eq} reads
\beq
    S(p) \= S_0(p) \+ S_0(p)\, \hat\Sigma  \,S(p).
\eeq
Solving for the self-energy, one obtains
\beq
    \hat\Sigma \= S^{-1}_0(p) \;-\; S^{-1}(p),
\eeq 
where 
\beq
S_0^{-1}(p) \= \matr{\psl + \hat \mu\gamma^0 - \hat m}{0}
                    {0}{\psl - \hat \mu\gamma^0 - \hat m}
\eeq
is the inverse bare quark propagator, while 
$S^{-1}$ is the inverse dressed propagator defined in \eq{SinvNG}.
Thus,
\beq
    \hat\Sigma \= \matr{0}
    {-\hspace{-3mm}\sum\limits_{A=2,5,7}\Delta_A\gamma_5\tau_A\lambda_A}
    {\sum\limits_{A=2,5,7}\Delta_A^*\gamma_5\tau_A\lambda_A}{0}.
\eeq
On the other hand, $\hat\Sigma$ can be evaluated diagrammatically. 
In Hartree approximation, it corresponds to the quark loop in 
Fig.~\ref{fig:dyson_eq} and is given by
\beq
\hat\Sigma 
\= 4iH\,\Gamma_i\intk\, \TrNG\left[\bar\Gamma_i  S(k)\right].
\eeq   
Comparing the two expressions for $\hat\Sigma$, we can read off the
following gap equations:
\beq
    \Delta_A \= 4H\intk\, \TrNG\left[\Gamma_{AA}^{s\downarrow}S(k)\right].
\label{gapDNG}
\eeq
We also see that the contributions of scalar vertices with $A\neq A'$
and of pseudoscalar vertices should vanish to be consistent. 
Using the explicit expression for the dressed propagator (see
Appendix \ref{appprop}), it can be shown that this is indeed the case.

\section{Chiral Ward-Takahashi identity}
\label{appwti}

In this appendix, we demonstrate that the dressed vertex functions and the
dressed quark propagator are consistent with the chiral Ward-Takahashi 
identity ($\chi$WTI) in the sense that, {\it if}\,  the $\chi$WTI holds, 
we recover the gap equation.
As in Sect.~\ref{eqmass}, we restrict ourselves to the case of equal
quark masses.

From \eqs{BSEANG} and (\ref{BSEPNG}) we obtain
\begin{alignat}{1}
    q_\mu\,\Gamma^\mu_{5j}&(p;q) \+ 2mi\,\Gamma_{5,j}(p;q)
\nonumber\\
    \,=\;&\Big((\qsl-2m)\gamma_5\,t_j\Big)_\mathit{NG}
\nonumber\\
    &\+ 4iH\;\Gamma_i \intk\, \TrNG\Big[\bar\Gamma_i\,  S(k+q)\,
\nonumber\\
&\hspace{10mm} \times\;
\Big(q_\mu\Gamma^\mu_{5j}(k;q) \,+\, 2mi\,\Gamma_{5,j}(k;q)\Big)
\,S(k)\Big].
\end{alignat}
Hence, imposing the $\chi$WTI, \eq{WTING}, one gets
\begin{alignat}{1}
    S^{-1}&(p+q)\,(\gamma_5\,t_j)_\mathit{NG} \,+\, 
    (\gamma_5\,t_j)_\mathit{NG}\,S^{-1}(p).
\nonumber\\
    \,=\;&\Big((\qsl-2m)\gamma_5\,t_j\Big)_\mathit{NG}
\nonumber\\
    &\+ 4iH\;\Gamma_i \intk\, \TrNG\Big[\Big(
      \bar\Gamma_i\,(\gamma_5\,t_j)_\mathit{NG} 
\nonumber\\
&\hspace{35mm} \,+\, (\gamma_5\,t_j)_\mathit{NG}\,\bar\Gamma_i\Big)\,S(k)\Big].
\end{alignat}
Using \eq{SinvNG} with equal masses and gap parameters,
one finds that the diagonal Nambu-Gorkov components are equal to the
first term on the r.h.s., and one is left with
\begin{alignat}{1}
    &\matr{0}{\hspace{-3mm}\Delta\hspace{-2mm}\sum\limits_{A=2,5,7}
      (\tau_A t_j^T + t_j\tau_A)\lambda_A}
      {-\Delta^*\hspace{-2mm}\sum\limits_{A=2,5,7}
      (\tau_A t_j + t_j^T\tau_A)\lambda_A}{0}
\nonumber \\
    &\= 4iH\;\Gamma_i \intk\, \TrNG\Big[\Big(
      \bar\Gamma_i\,(\gamma_5\,t_j)_\mathit{NG} 
\nonumber\\
&\hspace{37mm} \,+\, (\gamma_5\,t_j)_\mathit{NG}\,\bar\Gamma_i\Big)\,S(k)\Big].
\label{eqx}
\end{alignat}
Next we compute the sums over $A$ on the l.h.s. for any of the flavor
operators $t_j$.
For instance, for $t_j = t_{\pi^+} = \frac{\tau_1 + i\tau_2}{2\sqrt{2}}$ we 
find that the l.h.s. is equal to 
$\frac{1}{\sqrt{2}}(\Delta\,\Gamma_{57}^{p\uparrow} 
\+ \Delta^*\,\Gamma_{75}^{p\downarrow})$.
We conclude that the r.h.s. must vanish for all $\Gamma_i$, except for
$\Gamma_i = \Gamma_{57}^{p\uparrow}$ and 
$\Gamma_i = \Gamma_{75}^{p\downarrow}$.
In the first case, we have
$\bar\Gamma_i\,(\gamma_5\,t_{\pi^+})_\mathit{NG} 
\,+\, (\gamma_5\,t_{\pi^+})_\mathit{NG} \,\bar\Gamma_i \= 
-\frac{i}{\sqrt{2}}\Gamma_{77}^{s\downarrow}$. Thus by comparison with  the l.h.s.
we obtain
\beq
\Delta \= 4H\;\intk\, \TrNG\Big[\Gamma_{77}^{s\downarrow}
\,S(k)\Big],
\eeq
in agreement with one of the gap equations (\ref{gapDNG}) for the case of 
equal masses. 
The two other equations can be derived analogously, if we evaluate
\eq{eqx} for $t_j = t_{K^+}$ or $t_j = t_{K^0}$.
Moreover, the fact that most $\Gamma_i$ must not contribute to the
r.h.s. for a given $t_j$ can be used to show that 
scalar operators with $A \neq A'$ and pseudoscalar operators
do not contribute to the gap equation.

\section{Dressed quark propagator}
\label{appprop}

\subsection{General case}

The dressed quark propagator $S(p)$ is the inverse of the inverse quark 
propagator, defined in \eq{SinvNG}.
Following standard methods (see, e.g., 
Refs.~\cite{SRP,Ruster:2005jc,Blaschke:2005uj,Abuki:2005ms}), we write
\beq
    S^{-1}(p) \;\equiv\;  S^{-1}(p^0,\vec p) \=
    \gamma^0\,\Big(p^0 - A(\vec p)\Big),
\label{SinvA}
\eeq 
where $A(\vec p)$ is a hermitian $72 \times 72$ matrix, which does not 
depend on $p^0$.
Thus $A$ can always be diagonalized, i.e., we can find a unitary matrix 
$U(\vec p)$, so that 
\beq
    A(\vec p) \= U^\dagger(\vec p)\,D(\vec p)\,U(\vec p),
\eeq
with
\beq
    D(\vec p) \= \left( \begin{array}{ccc} 
                  \varepsilon_1(\vec p) &         & 0                      \\
                                        &  \ddots &                        \\
                   0                    &         & \varepsilon_{72}(\vec p) 
                  \end{array} \right)
\label{Adiag}
\eeq
being a diagonal matrix with eigenvalues 
$\varepsilon_1, \dots  \varepsilon_{72}$.
It can be shown that all eigenvalues are two-fold degenerate, and
for each eigenvalue $\varepsilon_i$, there is a counterpart $-\varepsilon_i$ 
in the spectrum. This means, there are basically 18 independent eigenvalues.
Moreover, part of the diagonalization is trivial because the matrix $A$ can 
be brought into block diagonal form by reordering of lines and columns. 
The remaining blocks are in general diagonalized numerically. 

Combining \eqs{SinvA} and (\ref{Adiag}),
the propagator is finally given by
\beq
    S(p) \=  U^\dagger(\vec p)\,
          \left( \begin{array}{ccc} 
           \frac{1}{p^0-\varepsilon_1(\vec p)} &  & 0  \\
           &  \ddots &    \\
           0 & & \frac{1}{p^0-\varepsilon_{72}(\vec p)} 
           \end{array} \right)
          \,U(\vec p)\,\gamma^0.
\eeq

\subsection{Equal quark masses}
\label{propex}
In the limit of an exact $SU(3)$ symmetry, 
we can give a closed expression for the quark propagator.
Straight forward inversion of \eq{SinvNG} for $m_u = m_d = m_s =m$ and 
$\Delta_{22} = \Delta_{55} = \Delta_{77} = \Delta$ yields
\begin{equation}
S = \left(\begin{array}{cc} S_{11} & S_{12} \\ S_{21} & S_{22}
  \end{array}\right),
\end{equation}
with 
\beq
S_{21} =  \Delta^*\,\frac{\pslm + m}{x_-}\;\gamma_5 \sum_{A = 2,5,7} \tau_A
\lambda_A\; S_{11}
\eeq
and
\beq
S_{11} =  \Big[\pslp - m \;-\; \frac{|\Delta|^2}{x_-}\,(\pslm - m) 
\hspace{-4mm} \sum_{A,A' = 2,5,7}\hspace{-4mm}\tau_A\tau_{A'}\,
\lambda_A\lambda_{A'}\Big]^{-1},
\eeq
where we have introduced the notations 
\beq
    p_{\pm} = p \pm \mu \gamma^0, \quad x_\pm =  p_{\pm}^2 - m^2.
\eeq
$S_{22}$ and $S_{12}$ are obtained from $S_{11}$ and $S_{21}$,
respectively, under the exchange $\mu \leftrightarrow -\mu$ and
$\Delta \leftrightarrow -\Delta^*$. 

The matrices $S_{ij}$ are $36\times$36 matrices representing 
the normal ($i=j$) and anomalous ($i\neq j$) Nambu-Gorkov components of $S$.
$S_{11}$ can explicitly be written as
\beq
    S_{11} \= S_- \+ \frac{1}{6} (T_- - S_-) \sum_{a=0}^8\tau_a\,\lambda_a,
\eeq
with 
\beq
    S_\pm \= \frac{x_\pm\,(\psl_\mp + m) \,-\, |\Delta|^2\,(\psl_\pm + m)}
             {(p_0^2 - E_8^{-\,2})(p_0^2 - E_8^{+\,2})},
\eeq
corresponding to the eigenvalue $\Delta$ of the gap matrix,
and
\beq
    T_\pm \= \frac{x_\pm\,(\psl_\mp + m) \,-\, 4|\Delta|^2\,(\psl_\pm + m)}
             {(p_0^2 - E_1^{-\,2})(p_0^2 - E_1^{+\,2})}
\eeq
corresponding to the eigenvalue $2\,\Delta$ of the gap matrix.
The octet and singlet dispersion relations for particles ($-$) and
antiparticles ($+$) are given by
\beq
    E_8^\mp \= \sqrt{(\sqrt{\vec p^{\,2} + m^2} \mp \mu)^2 + |\Delta|^2}
\eeq
and
\beq
    E_1^\mp \= \sqrt{(\sqrt{\vec p^{\,2} + m^2} \mp \mu)^2 + 4|\Delta|^2},
\eeq
respectively.

\section{Pion decay constant in the chiral limit}
\label{appfpi}

In this appendix we derive a semi-analytical
expression for the pion
decay constant in the chiral limit, which is used in 
Sect.~\ref{sec:fpinum} to discuss the deviations from the weak coupling limit.
The $\eta'$ decay constant can be obtained in a similar way but we do not
discuss this here.

Starting point is \eq{fieqm} for the (time-like)
decay constant in the case of equal quark masses. 
In the chiral limit, we have to evaluate this formula at $m_i=0$,
Since both sides of this equation vanish in this limit, this means
that the decay constant is given by the derivative of the r.h.s. with 
respect to $q_0$. 
Moreover, we can employ the ``Goldberger-Treiman relation'', \eq{GTNG}, 
to eliminate the coupling constant $g_i$. We then obtain
\begin{alignat}{1}
    f_i^2 = -\Delta\,\Big(\frac{d}{dq_0}
             &\intk 
\nonumber\\    
    &\;\TrNG[\bar\Gamma_i'(q)\,  S(k+q)
    (\gamma^0\gamma_5\,t_i)_\mathit{NG} S(k)]\Big)\!\Big|_{q = 0}.
\label{fpiint}
\end{alignat}
Here we have indicated explicitly that the vertex function $\Gamma_i'$
depends on the momentum (see \eq{Gammapiq}) and is therefore subject
to the derivative. 
We may thus write
\beq
    f_i^2 = \tilde f_i^2 + \delta f_i^2, 
\eeq
where
\begin{alignat}{1}
    &\tilde f_i^2 =
\nonumber\\    
&-\Delta \intk \;\TrNG[\bar\Gamma_i'(0)\,  \frac{dS(k+q)}{dq_0}\Big|_{q = 0}
    (\gamma^0\gamma_5\,t_i)_\mathit{NG}\, S(k)]
\label{fpiintappa}
\end{alignat}
corresponds to the contribution where the derivative acts on the propagator,
while
\begin{alignat}{1}
   &\delta f_i^2 = 
\nonumber\\ 
&-\Delta \intk \;\TrNG[\frac{d \bar\Gamma_i'(q)}{dq_0}\Big|_{q=0} S(k)\,
    (\gamma^0\gamma_5\,t_i)_\mathit{NG}\, S(k)]
\label{fpiintappb}
\end{alignat}
corresponds to the contribution where the derivative acts on the vertex 
function.

The evaluation of $\tilde f_i^2$ is tedious, but straight forward.
Inserting $\Gamma_i'(0) = \Gamma_i'\hspace{-0.5mm}^{(0)}$ 
from \eqs{eqmops} or (\ref{eqmopshf}),
respectively, $t_i$ from \tab{tab:qnop1}
as well as the expressions for the quark propagator given in
appendix~\ref{propex} in the chiral limit,
the result for the octet mesons (``pions'') reads
\begin{alignat}{1}
\tilde{f}_\pi^2 \= & \frac{\mu^2}{216 \pi^2} \nonumber \\
 \Big\{ & \frac{2}{y^2}\,\Big[(\alpha_1^+ - \alpha_8^+)(x+1)(3x^2 - 2x + 1) 
\nonumber \\
& + (\alpha_1^- - \alpha_8^-)(x-1)(3x^2 + 2x + 1)\Big] \nonumber \\
& -  \alpha_8^+ (9x-31) - \alpha_8^- (9x+31) \nonumber\\
& -  16\alpha_1^+ + 16\alpha_1^- - 45\beta \nonumber \\
& + 24\,\Big[\ln(\alpha_1^+ + x+1) + \ln(\alpha_1^- + x-1)\nonumber \\
& \hspace{6mm}- \ln(\alpha_8^+ + x+1) - \ln(\alpha_8^- + x-1)\Big] \nonumber \\
& + 45\,\Big[\frac{x+1}{\alpha_8^+} + \frac{x-1}{\alpha_8^-} + \frac{1}{\beta}\Big]\nonumber \\
& - 48\ln 2 \;-\; 54\,y^2\ln y^2 \nonumber \\
& + 54y^2\,\Big[\ln(\alpha_8^+ + x+1) + \ln(\alpha_8^- + x-1)\Big]\nonumber \\
& - 45y^2\,\Big[\frac{x-1}{\alpha_8^+} + \frac{x+1}{\alpha_8^-} - \frac{1}{\beta}
\Big] \;\Big\},
\label{fjeq}
\end{alignat}
where the 3-momentum integral has been regularized by a cutoff $\Lambda$
(as in the numerical calculations)
and we have introduced the abbreviations
\beq
 \alpha_1^{\pm} = \sqrt{(x\pm 1)^2 + 4\, y^2}, \quad
 \alpha_8^{\pm} = \sqrt{(x\pm 1)^2 + y^2}, 
\label{alphai}
\eeq
and
\beq
 \beta = \sqrt{1 + y^2}.
\eeq
Moreover, $x = \frac{\Lambda}{\mu}$ and $y = \frac{\Delta}{\mu}$, as
defined in \eq{xydef}.

The evaluation of $\delta f_i^2$ is more difficult because we need to
know the derivative of the vertex function. For the flavored mesons
this is encoded in the mixing angle $\varphi(q)$, cf.~\eq{Gammapiq}.
We can therefore write
\beq
    \frac{d\bar\Gamma_i'(q)}{dq_0}\Big|_{q=0} \=
    \frac{d\varphi(q)}{dq_0}\Big|_{q=0} \bar\Gamma'^\perp_i,
\eeq
where
\beq
    \Gamma'^\perp_i \=
    \frac{\partial\Gamma_i'}{\partial\varphi}
    \Big|_{\varphi = \frac{\pi}{4}}
\eeq
is the vertex of the orthogonal state with the same quantum numbers
as the meson $i$ (e.g., 
$\Gamma'^\perp_{\pi^+}= -\frac{i}{\sqrt{2}} (\Gamma_{57}^{p\uparrow} 
                              + \Gamma_{75}^{p\downarrow})$).
Hence
\begin{alignat}{1}
  &\delta f_i^2  =
\nonumber\\  
  &-\Delta \intk \;\TrNG[\bar\Gamma'^\perp_i\, S(k)\,
    (\gamma^0\gamma_5\,t_i)_\mathit{NG}\, S(k)] \frac{d\varphi(q)}{dq_0}\Big|_{q=0}.
\end{alignat}
Evaluating this part for the octet mesons, one finally obtains
\begin{alignat}{1}
\delta f_\pi^2 \=& 
\frac{\mu ^2}{108\pi^2} 
\nonumber\\ 
\Big\{
&- 108 y^2\, \ln y^2 -48 y^2 \,\ln 2
\nonumber \\ 
&+ 2(\alpha_1^+-\alpha_8^{+})(x^2 - x + 1) 
\nonumber \\ 
&- 2(\alpha_1^--\alpha_8^{-})(x^2 + x + 1) 
\nonumber \\ 
&+4y^2(\alpha_8^+ - \alpha_8^-) -16y^2(\alpha_1^+ -\alpha_1^-)
\nonumber \\ 
&+12 \, y^2 \
\Big [2 \ln (\alpha_1^-+x-1)+2 \ln \
(\alpha_1^++x+1)
\nonumber \\ 
&\hspace{10mm} +7 \ln (\alpha_8^-+x-1)+7 \ln \
(\alpha_8^++x+1) \Big ] 
\nonumber \\ 
&-45 y^2 \Big[\frac{x^2+4x+2}{\alpha_8^+}-\frac{x^2-4x+2}{\alpha_8^-}\Big ]
\nonumber \\ 
&-90 y^4 (\frac{1}{\alpha_8^+}-\frac{1}{\alpha_8^-})
\Big\}\,\mu\,\frac{d\varphi(q)}{dq_0}\Big|_{q=0}.
\label{fpimomentum}
\end{alignat}

One could also try to derive an analytical expression for 
the derivative $\frac{d\varphi(q)}{dq_0}|_{q=0}$
from the polarization-loop matrix \eq{Jij},
but this is beyond the scope of this paper.

\section{Exact formula for $A$}

The exact result of \eq{Aex} is given by
\begin{alignat}{1}
A = &\frac{\Delta^2}{384 \pi^2}
\nonumber\\
   \Big\{
           &-48 \ln{y^2} - 32 \ln{2}
           + 4(\alpha_1^- -\alpha_1^+) + 44(\alpha_8^- -\alpha_8^+)
\nonumber\\
 &+ (16+3y^2)\Big[ \ln(\alpha_1^- +x-1) + \ln(\alpha_1^+ +x+1) \Big]
\nonumber\\
 &+ (32-3y^2)\Big[ \ln(\alpha_8^- +x-1) + \ln(\alpha_8^+ +x+1) \Big]
\nonumber\\
 &- 6y^2 \ln{2}
\nonumber\\
 &+9 y^2 \frac{\beta_2}{\beta_3}\; \Big[ 
   \; \ln[4(x+1) + \alpha_1^+ \beta_2\beta_3 - (3x-13)y^2] 
\nonumber\\
 & \hspace{14.0mm}  
   -  \ln[4(x-1) - \alpha_1^- \beta_2\beta_3 - (3x+13)y^2] 
\nonumber\\
 & \hspace{14.0mm}  
   +  \ln[4(x+1) + \alpha_8^+ \beta_2\beta_3 + (3x+7)y^2]
\nonumber\\
 & \hspace{14.0mm}  
   -  \ln[4(x-1) - \alpha_8^- \beta_2\beta_3 + (3x-7)y^2]\;\;
\Big]
\Big\},
 \label{eq_A_exact}
\end{alignat}
with $\alpha_i^\pm$ as defined in \eq{alphai}, 
\beq
 \beta_2 = \sqrt{4 + y^2} , \quad  \beta_3 = \sqrt{4 + 9y^2},
\eeq
and $x = \frac{\Lambda}{\mu}$ and $y = \frac{\Delta}{\mu}$, as
defined in \eq{xydef}.

\end{document}